\documentclass[lettersize,journal]{IEEEtran}
\usepackage{amsmath,amsfonts}
\usepackage{array}
\usepackage[caption=false,font=normalsize,labelfont=sf,textfont=sf]{subfig}
\usepackage{textcomp}
\usepackage{stfloats}
\usepackage{url}
\usepackage{verbatim}
\usepackage{graphicx}
\usepackage{cite}
\usepackage{bm}
\usepackage{hyperref}
\usepackage[ruled]{algorithm2e}
\usepackage{tikz,xcolor} 

\begin{document}

\definecolor{lime}{HTML}{A6CE39}
\DeclareRobustCommand{\orcidicon}{
\begin{tikzpicture}
\draw[lime, fill=lime] (0,0)
circle[radius=0.13]
node[white]{{\fontfamily{qag}\selectfont \tiny \.{I}D}};
\end{tikzpicture}
\hspace{-2mm}
}
\foreach \x in {A, ..., Z}{%
\expandafter\xdef\csname orcid\x\endcsname{\noexpand\href{https://orcid.org/\csname orcidauthor\x\endcsname}{\noexpand\orcidicon}}
}
\newcommand{\orcidauthorA}{0000-0002-7296-2589}
\newcommand{\orcidauthorB}{0000-0002-1546-4364}
\newcommand{\orcidauthorD}{0000-0003-0907-4594}

\title{Quantum Federated Learning for Distributed \\Quantum Networks}

\author{Kai Yu\hspace{-1.5mm}\orcidA{}, Fei Gao\hspace{-1.5mm}\orcidB{}, and Song Lin\hspace{-1.5mm}\orcidD{}
\thanks{$\bullet$ Kai Yu is with the College of Computer and Cyber Security, Fujian Normal University, Fuzhou, 350117, China. E-mail: ykai95@163.com.}
\thanks{$\bullet$ Fei Gao, the corresponding author, is with the State Key Laboratory of Networking and Switching Technology, Beijing University of Posts and Telecommunications, Beijing, 100876, China. E-mail: gaof@bupt.edu.cn.}
\thanks{$\bullet$ Song Lin, the corresponding author, is with the College of Computer and Cyber Security, Fujian Normal University, Fuzhou, 350117, China. E-mail: lins95@gmail.com.}
}

\markboth{Journal of \LaTeX\ Class Files,~Vol.~14, No.~8, August~2021}%
{Shell \MakeLowercase{\textit{et al.}}: A Sample Article Using IEEEtran.cls for IEEE Journals}


\maketitle

\begin{abstract}
Federated learning is a framework for learning from distributed networks. It attempts to build a global model based on virtual fusion data without sharing the actual data. Nevertheless, the traditional federated learning process encounters two challenges: high computational cost and message transmission security. To address that, we propose a quantum federated learning for distributed quantum networks by utilizing quantum characteristics. First, we give two methods to extract the data information to the quantum state. It can cope with different acquisition frequencies of data information. Next, a quantum gradient descent algorithm is provided to help clients in the distributed quantum networks to train local models in parallel. Compared with the classical counterpart, the proposed algorithm achieves exponential acceleration in dataset scale and quadratic speedup in data dimensionality. And, a quantum secure multi-party computation protocol with Chinese residual theorem is designed. It could avoid errors and overflow problems that may occur in the process of large number operation. Security analysis shows that the protocol can resist common external and internal attacks. Finally, to demonstrate the effectiveness of the proposed framework, we use it to train a federated linear regression model and simulate the essential computation steps on the IBM Qiskit simulator.
\end{abstract}


\section{\label{sec:1}Introduction}
\IEEEPARstart{W}{ith} the development of information network, more and more data are generated and stored in distributed network system \cite{Jia2019}. Integrating data from distributed networks enables the extraction of valuable information. Nevertheless, a significant proportion of the data contains sensitive and private information, making data owners hesitant to share it \cite{Gu2022}. This situation has led to the emergence of federated learning (FL), which is a distributed machine learning (ML) method \cite{brendan2017}. FL improves data privacy by localizing data and training it without sharing the raw data. This not only makes effective use of distributed data resources, but also facilitates the development of information network technology. However, the volume of locally trained data can be huge. At this point, the computing power of traditional computers will face great challenges. Furthermore, the transmission of training results poses a threat to user privacy, as it can provide an opportunity for attackers to infer sensitive information. While some classical passwords have been used to safeguard communication security, development in hardware poses a persistent threat to their security.

\par Quantum information processing (QIP) is an emerging field that explores the interaction between quantum mechanics and information technology. It sustains to show its charm, attracting the attention of scholars. In 1984, Bennett and Brassard proposed the famous BB84 protocol \cite{bennett1984}, which perfectly achieves a key distribution task between two remote parties. Subsequently, scholars utilized quantum information processing to ensure information security and proposed a series of quantum cryptography protocols. In contrast to the security of classical cryptography protocols that are based on the assumption of computational complexity, these protocols' security relies on physical properties such as the Heisenberg uncertainty principle, which makes them unconditionally secure in theory. Quantum cryptography has developed as a significant application of quantum information processing, including quantum key distribution \cite{Gisin2002,scarani2009,schwonnek2021}, quantum secret sharing \cite{PhysRevA.59.1829, PhysRevA.59.162, PhysRevLett.83.648}, quantum secure direct communication \cite{kim2002, PhysRevA.68.042317}, and so on. Another exciting application of quantum information processing is quantum computing. It provided quantum speedup to certain classes of problems that are intractable on classical computers. For example, the factorization of large numbers via Shor algorithm \cite{Shor1999} can provide exponential speedup. Furthermore, quantum computing has also made some advances in machine learning, such as the quantum linear systems solving algorithms \cite{HHL,WanLC2021,LiuHL2023}, quantum regression \cite{YCH2021IEEE,chen2022}, quantum neural network \cite{Scala2023, Wuprl2023}, variational quantum algorithms (VQA) \cite{larose2019variational,LiuHL2021,ZhangSX2023}, and so on.

\par Motivated by the advantages shown by quantum cryptography and quantum computing respectively in improving transmission security and computing speed, scholars have attempted to utilize QIP to address the challenges faced by FL. In 2021, Li et al. focused on the security issue of FL \cite{li2021QFL}. They proposed a private single-party delegated training protocol based on blind quantum computing for a variational quantum classifier, and then extended the protocol to quantum FL combined with differential privacy. This protocol can exploit the computing advantage of remote quantum servers with privacy of sensitive data. In 2024, Ren et al. proposed a quantum FL to solve the privacy preservation issue for the smart cyber-physical grid dynamic security assessment problem \cite{Ren2024}. Moreover, Chen and Yoo proposed a quantum FL scheme with a hybrid quantum-classical machine learning model, focusing on improving the efficiency of the local training \cite{chen2021QFL}. In their way, the classical convolutional network is used to extract data features and compress them into vectors which are input into variable quantum circuits for training. Compared with the classical process, this method can achieve the same level of accuracy more quickly. In 2022, Huang et al. utilized a variational quantum algorithm to estimate the gradient of the local model to avoid analyzing the gradient too costly \cite{huang2022QFL}. As variational quantum algorithms approximate the target results by using circuits with variables, they are different from quantum algorithms that calculate the target results through the evolution of quantum gates. Therefore, we further explore the realization of FL with quantum resources.

\par In this paper, we focus on the quantum algorithm running on ordinary quantum computers and present a quantum federated learning based on gradient descent (QFLGD). It aims to provide a unified, secure, and effective gradient distribution estimation scheme with distributed quantum networks. In QFLGD, we propose two data preparation methods by analyzing the different acquisition frequencies of static data (the local training data) and dynamic data (the parameters that need to be updated during iteration). That can reduce the requirement of QFLGD on the performance of quantum random memory. At the same time, two main processes of FL are implemented in QFLGD, which exploit quantum properties. The first one is a quantum gradient descent (QGD) algorithm. It facilitates the acceleration of the training gradient for the client. QGD provides the client with a classical gradient at each iteration, which can be directly used to learn classical model parameters. Compared with the classical counterpart, this quantum process has exponential acceleration in terms of data scale and quadratic speedup in data dimensionality. The other is a quantum secure multi-party computation (QSMC) protocol, which allows the aggregation of gradients to securely be done with quantum communication networks. That is, the server is able to calculate the federated gradients without the client sharing the local gradients. Furthermore, the application of the Chinese remainder theorem in QSMC makes it possible to avoid errors and overflow problems that may occur during the calculation of large numbers. The proposed quantum federated learning framework can improve the local computing efficiency and data privacy of FL. We also apply QFLGD to train the federated linear regression (FLR) and give its numerical experiment to verify the correctness.

\par The remainder of this paper is organized as follows. The classical FL is reviewed in Sec. \ref{sec:2}. In Sec. \ref{sec:3}, we propose the framework for QFLGD. In Sec. \ref{sec:4}, we analyze the time complexity and the security of QFLGD. Furthermore, an application to train the FLR and the numerical experiment are shown in Sec. \ref{sec:5}. In Sec. \ref{sec:6}, we give the conclusion of our work.

\section{\label{sec:2}Review of classical FL} 

\par
\begin{figure}
        \centering
        \includegraphics[width=0.5\textwidth]{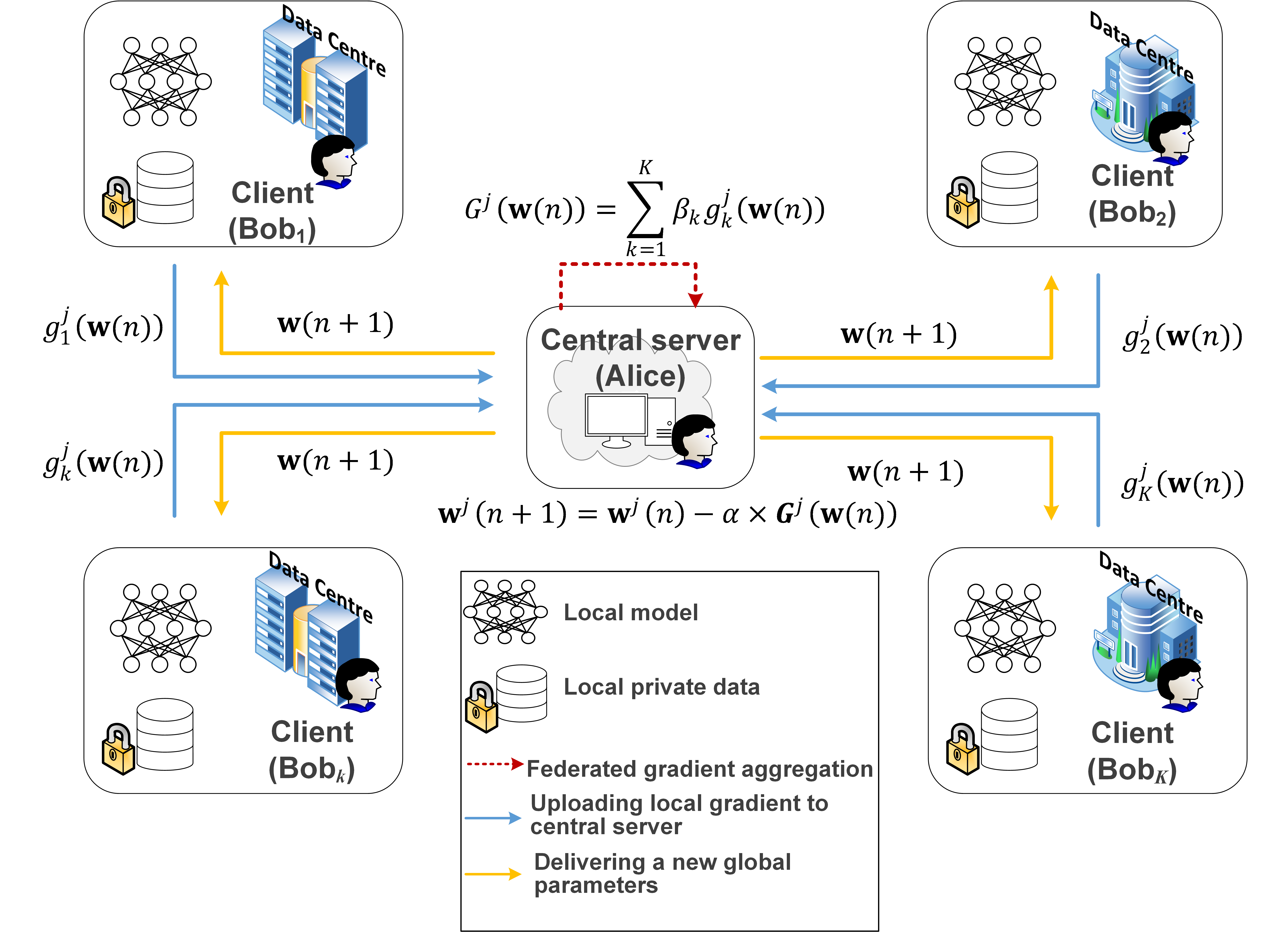}
        \caption{Schematic diagram of the federated learning based on gradient descent. $\mathbf{w}^{j}(n)$ is denoted the $j$th element of the parameter vector $ \mathbf{w}(n)$. $\bm{G}^{j} \left( \mathbf{w}(n) \right)$ is the $j$th component of the global gradient. $\beta_{k}$ is the aggregation weight. $\bm{g}^{j}_{k}\left( \mathbf{w}(n) \right)$ is represented as the $j$th element of local gradient of the client ${\rm Bob}_{k}$. $\alpha$ is a learning rate.}
    \label{fig:1}
    \end{figure}

 To clarify the framework of QFLGD in the distributed quantum networks, this section offers a overview of the fundamental ideas and processes of traditional FL. FL is a collaborative ML approach in which multiple clients train a shared model without exchanging raw data. A popular learning framework is FL based on gradient descent \cite{brendan2017}, which is depicted in Fig. \ref{fig:1}. It mainly includes the following parts.
\par \textit{A) Data preparation and model initialization.} In the FL framework, data is derived from various clients in a distributed network, such as hospital medical information, preference options in business surveys, and other sensitive data \cite{TPDSWang2019}. We consider general federated learning with $K$ clients participating in the model training. The server (Alice) initializes a global model that requires training parameters $\mathbf{w} = \left( \omega_{0},\omega_{1},\cdots,\omega_{D-1} \right)$ to make it more efficient. And the server distributes it to clients. The client ${\rm Bob}_{k}$ ($k = 1,2,\cdots,K$) collects and preprocesses $M_{k}$ data samples $\left( {\mathbf{x}_{0},y_{0}} \right),\left( {\mathbf{x}_{1},y_{1}} \right),\cdots,\left( {\mathbf{x}_{M_{k}-1},y_{M_{k}-1}} \right)$, where $\mathbf{x}_{i} \in \mathbb{R}^{D}$ and $y_{i}$ is the corresponding label.

\par \textit{B) Local training.} To train the model, clients use standard ML algorithms without sharing raw data. The trained ML model evaluation task is expressed as minimizing the cost function, such as minimizing mean square error (MSE) loss function
     \begin{equation}
        {\min\limits_{\mathbf{w}}E} = \frac{1}{2M_{k}}{\sum\limits_{i = 0}^{M_{k}-1}\left\lbrack {f\left( {\mathbf{x}_{i} \cdot \mathbf{w}} \right) - y_{i}} \right\rbrack^{2}},
     \label{eq:1}
     \end{equation}
where $f$ is the activation function. This function is usually expressed as minimizing the difference between the model output and the expected output. In this case, the model optimization is to find the gradient of $E$ with respect to $\mathbf{w}$ to adjust the model parameters. For the client ${\rm Bob}_{k}$ ($k = 1,2,\cdots,K$), he can obtain
     \begin{equation}
        \bm{g}_{k}^{j}\left( \mathbf{w} \right) = \frac{1}{M_{k}}{\sum\limits_{i = 0}^{M_{k}-1}{F\left( {\mathbf{x}_{i} \cdot \mathbf{w}} \right)}}\mathbf{x}_{i}^{j},\quad j = 0,1,\cdots,D-1,
     \label{eq:2}
     \end{equation}
with his data. Here, $F\left( {\mathbf{x}_{i} \cdot \mathbf{w}} \right)$ is represented as $\frac{\partial(f({\mathbf{x}_{i} \cdot \mathbf{w}}))}{\partial({\mathbf{x}_{i} \cdot \mathbf{w}})} \left\lbrack {f\left( {\mathbf{x}_{i} \cdot \mathbf{w}} \right) - y_{i}} \right\rbrack $, $\bm{g}_{k}^{j}\left( \mathbf{w} \right)$ is denoted the $j$th element of the local gradient $\bm{g}_{k}\left( \mathbf{w} \right)$, and $\mathbf{x}_{i}^{j}$ is labeled the $j$th element of the sample $\mathbf{x}_{i}$.

\par \textit{C) Model aggregation and update.} The server (Alice) collects the gradients trained by all clients, and calculates the federated gradient
     \begin{equation}
        \bm{G}^{j} \left( \mathbf{w} \right) = {\sum\limits_{k = 1}^{K}{\beta_{k}}{\bm{g}_{k}^{j}\left( \mathbf{w} \right)}},
     \label{eq:5}
     \end{equation}
where ${\beta_{k}}={M_{k}}/ ({\sum_{k=1}^{K}{M_{k}}})$ and $\bm{G}^{j}\left( \mathbf{w} \right)$ is represented as the $j$th element of the federated gradient $\bm{G}\left( \mathbf{w} \right)$. Then, Alice updates the global model parameters. Specifically, she adjusts the parameter to
     \begin{equation}
        \mathbf{w}^{j}\left( {n + 1} \right) = \mathbf{w}^{j}(n) - \alpha \times \bm{G}^{j}\left( {\mathbf{w}(n)} \right),
     \label{eq:6}
     \end{equation}
for $j = 0,1,\cdots,D-1$, in the $n$th iteration. In Eq.(\ref{eq:6}), $\alpha$ is a learning rate \cite{Xue2024}.

\par \textit{D) Model evaluation and distribution.} The server (Alice) evaluates the performance of the global model and sends the global model parameters to the clients for further local training if it has not yet converged (i.e., ${\sum\limits_{j = 0}^{D-1}\left[ {\bm{G}^{j}\left( \mathbf{w}(n) \right)} \right]^{2}} > \varepsilon$ where $\varepsilon$ is a threshold about gradient). Once the condition is satisfied, Alice announces that the training stops and distributes the model.

\par It is notable that the time consumption of FL is mainly in the calculation of the local gradient. For the client, he computes a gradient element takes $O(D)$ time to estimate the inner product ${\mathbf{x}_{i} \cdot \mathbf{w}}$ and $O(M)$ time to sum. In general, it takes $D$ repetitions to estimate all elements of the local gradient. Therefore, ${\rm Bob}_k$ takes $O(MD^2)$ time to calculate all the elements of the local gradient on a classical computer. In the era of big data, this is surely a very expensive calculation. Moreover, the security of federation learning may be compromised during local gradient aggregation. Traditional encryption methods can improve the security of this process. However, with the development of quantum technology, there are threats to traditional encryption methods.
%
%
\section{\label{sec:3}Quantum Federated Learning based on Gradient Descent Algorithm}

\par
\begin{figure}[!t]
        \centering
        \includegraphics[width=0.48\textwidth]{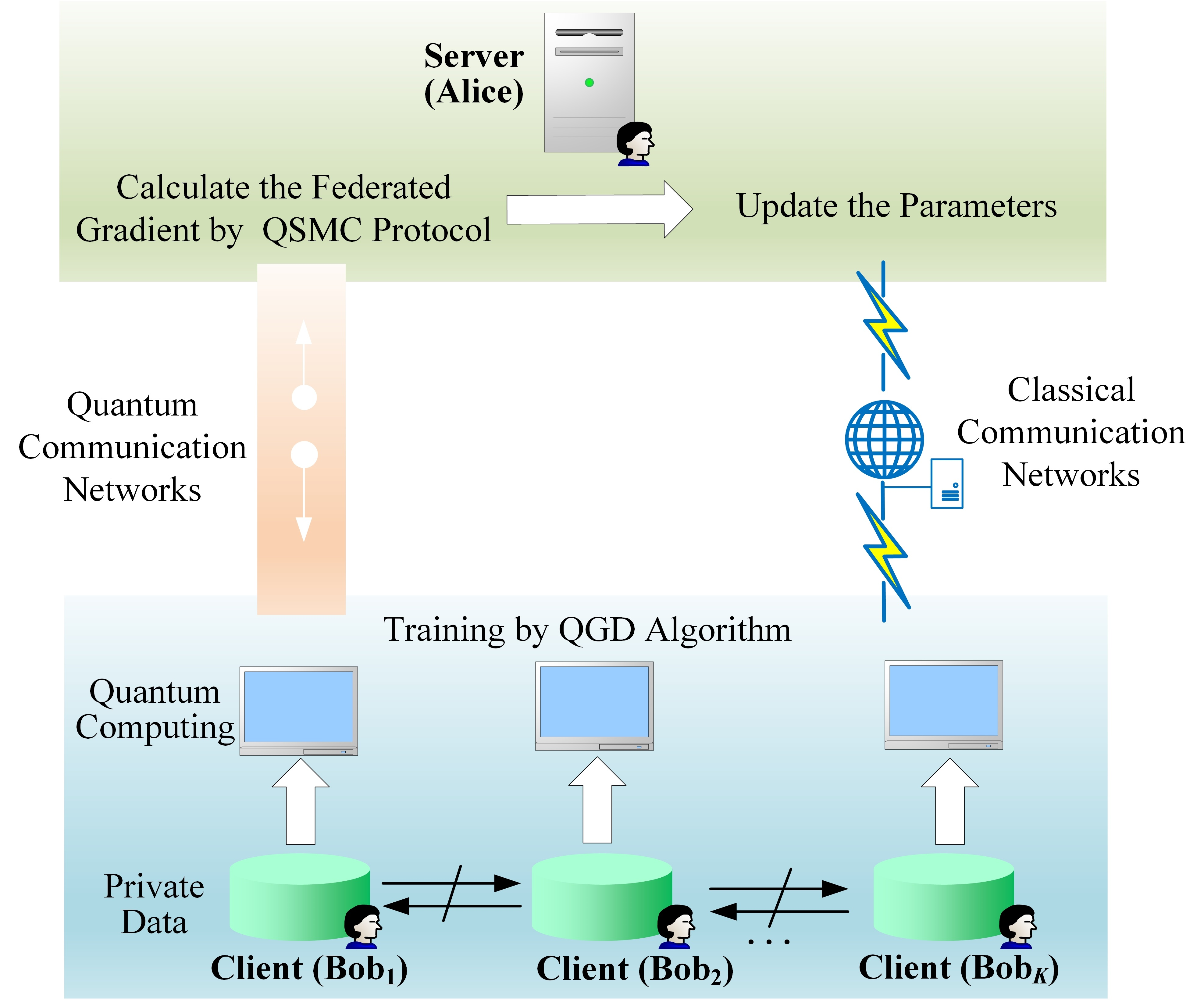}
        \caption{Schematic illustration of QFLGD.}
    \label{fig:QFLGDM}
    \end{figure}

 In this section, we present the QFLGD, which focuses on the parallel and private computing architectures for data in distributed quantum networks. This distributed quantum network typically consists of a server and several clients with quantum computing capabilities. We first give ways to extract the data information to the quantum state. Subsequently, we propose a QGD algorithm that clients use to estimate the gradient locally. A QSMC protocol is designed to perform a private calculation of the global gradient when the server aggregates the training results of clients. Finally, the server updates the global parameters and shares the results with clients. The schematic diagram of QFLGD framework is presented in Fig. \ref{fig:QFLGDM}.

\subsection{\label{sec:3.0}Quantum data preparation and model initialization}

\par Similar to the classical FL, a dataset $\mathbf{X} = \left[ \mathbf{x}_{0},\mathbf{x}_{1},\cdots,\mathbf{x}_{M-1} \right]$ is chosen by a client in quantum FL, where $\mathbf{x}_{i} \in \mathbb{R}^{D}$. $\mathbf{y} = \left( y_{0},y_{1},\cdots,y_{M-1} \right)$ is the corresponding label of each sample of $\mathbf{X}$, respectively. For convenience, assuming that $D=2^{L}$ for some $L$; otherwise, some zeros are inserted into the vector. Furthermore, the server initializes an ML model with parameters. The learnable parameters of the model are represented by a vector $\mathbf{W} \in \mathbb{R}^{D}$, which can be optimized using gradient descent. The ability of quantum computers to effectively solve practical problems depends on encoding this information into quantum states as input to a quantum algorithm. Here, we give methods to extract the data and parameters information to quantum states.

\par Considering the quantum oracles
     \begin{equation}
        O_{X}:\left|  i \right\rangle \left|  j \right\rangle \left| 0 \right\rangle \longrightarrow\left|  i \right\rangle \left|j \right\rangle| \mathbf{x}_{i}^{j}\rangle,
     \label{eq:7}
     \end{equation}
and
     \begin{equation}
         O_{y}:\left|  i \right\rangle \left|  0 \right\rangle \longrightarrow\left| i \right\rangle  |y_{i}\rangle,
     \label{eq:8}
     \end{equation}
are provided, where $\mathbf{x}_{i}^{j}$ represents the $j$th element of the $i$th vector of the data set $\mathbf{X}$. These two oracles can respectively access the entries of $\mathbf{x}_{i}$, $\mathbf{y}$ in time $O(\text{polylog}(MD))$ and $O(\text{polylog}(M))$ \cite{YCH2021IEEE,wossnig2018}, when the data are stored in quantum random access memory (QRAM) \cite{giovannetti2008} with an appropriate data structure \cite{kerenidis2016}. In addition, the operation
     \begin{equation}
         U_{nf}:\left| i \right\rangle \left| 0 \right\rangle \longrightarrow\left| i \right\rangle \left| \left\| \mathbf{x}_{i} \right\| \right\rangle.
     \label{eq:9}
     \end{equation}
is required, which could access the $2$-norm of the vector $\mathbf{x}_{i}$. Inspired by Ref. \cite{mitarai2019}, $U_{nf}$ can be implemented in time $O\left( {{\text{polylog}(D)}/\epsilon_{m}} \right)$ employing controlled rotation \cite{cong2016} and quantum phase estimation (QPE) \cite{brassard2002}. The details are shown in appendix \ref{appendix A}. According to these assumptions, the processes of the quantum data preparation are described as follows.

\par $(A1)$ In this step, the data information is extracted to the state $\left| {\phi\left( \mathbf{x}_{i} \right)} \right\rangle$. Firstly, three quantum registers are prepared in state $|i\rangle_{1} |0^{\otimes{\log{D}}}\rangle_{2} |0^{\otimes{q}}\rangle_{3} $, where the subscript numbers denote different registers. The $q$ is labeled as the qubits that are enough to store the information about the elements of data, i.e., $2^{q} - 1 > {\max\limits_{i,j}|\mathbf{x}_{i}^{j}|}$. After that, $H^{\otimes{\log{D}}}$ is applied on the second register to generate a state
     \begin{equation}
         \frac{1}{\sqrt{D}}{\sum\limits_{j = 0}^{D-1}\left|  i \right\rangle_{1} } |j\rangle_{2} | 0^{\otimes q} \rangle_{3}.
     \label{eq:10}
     \end{equation}

\par Secondly, the quantum oracle $O_X$ is performed on the three registers. These registers are in a state
     \begin{equation}
         \frac{1}{\sqrt{D}}{\sum\limits_{j = 0}^{D-1} |i\rangle_{1} } |j\rangle_{2} |\mathbf{x}_{i}^{j}\rangle_{3}.
     \label{eq:11}
     \end{equation}
Subsequently, a qubit in the state $|0\rangle$ is added and rotated to $\sqrt{1 - ({c_{1}\mathbf{x}_{i}^{j}})^{2}}\left|0 \right\rangle + c_{1}\mathbf{x}_{i}^{j}\left|1 \right\rangle$ controlled on $|\mathbf{x}_{i}^{j} \rangle$, where $c_{1} = {1/ {\max\limits_{i,j} |\mathbf{x}_{i}^{j}|} }$. The system becomes
     \begin{equation}
         \frac{1}{\sqrt{D}}{\sum\limits_{j = 0}^{D-1} |i\rangle_{1} } |j\rangle_{2} |\mathbf{x}_{i}^{j} \rangle_{3} \left[\sqrt{1 - ({c_{1}\mathbf{x}_{i}^{j}})^{2}} |0\rangle + c_{1}\mathbf{x}_{i}^{j}|1\rangle \right]_{4}.
     \label{eq:12}
     \end{equation}

\par Finally, the inverse $O_X$ operation is applied on the third register. The quantum state
     \begin{equation}
         |i\rangle | {\phi\left( \mathbf{x}_{i} \right)} \rangle = |i\rangle_{1} \frac{1}{\sqrt{D}} {\sum\limits_{j = 0}^{D-1} |j\rangle_{2} \left[\sqrt{1 - ({c_{1}\mathbf{x}_{i}^{j}})^{2}} |0\rangle + c_{1}\mathbf{x}_{i}^{j}|1\rangle \right]_{4}},
     \label{eq:13}
     \end{equation}
could be obtained via discarding the third register. The process is denoted as $U_{\mathbf{x}_i}$, which generates the state $\left| {\phi\left( \mathbf{x}_{i} \right)} \right\rangle$ in time $O(\text{polylog}(D)+q)$.

\par $(A2)$ In order to train the gradient, the parameter $\mathbf{w}(n)$ should be introduced in the $n$th iteration. Thus, it is necessary to generate a quantum state, which contains the information of $\mathbf{w}(n)$. Depending on the fact that the parameter is different in each iteration, there are two methods to prepare the quantum state.

\par One way is based on the assumption that QRAM is allowed to read and write frequently. For the information of $\mathbf{w}^{j}(n)$ ($j=0,1,\cdots,D-1$) are written in QRAM timely, the quantum state
     \begin{equation}
         \frac{1}{\sqrt{D}} \sum\limits_{j = 0}^{D-1} |j\rangle \left[\sqrt{1 - ({c_{2}\mathbf{w}^{j}(n)})^{2}}\left|0 \right\rangle + c_{2}\mathbf{w}^{j}(n)|1\rangle \right]
     \label{eq:14}
     \end{equation}
can be produced by the processes similar to step $(A1)$ with the help of the oracle $O_{\mathbf{w}}$ ($O_{\mathbf{w}}\left| j \right\rangle \left| 0 \right\rangle \longrightarrow\left| j \right\rangle |  {\mathbf{w}^{j}(n)} \rangle $), where $c_{2} = {1/\left\| \mathbf{w}(n) \right\|}$ and $\mathbf{w}^{j}(n)$ is denoted as the $j$th element of the parameter vector in the $n$th iteration. This way can be implemented in time $O(\text{polylog}(D)+q)$.

\par For another, the parameter is extracted into the quantum state based on the operation $R(\vartheta)=\cos{(\vartheta)} |0\rangle\langle0| - \sin{(\vartheta)} |0\rangle\langle1| + \sin{(\vartheta)} |1\rangle\langle0| + \cos{(\vartheta)} |1\rangle\langle1|$, which is inspired by Ref. \cite{shao2019fast}. In this way, the $\mathbf{w}(n)$ is not required to be written in QRAM. The following are described as the processes.

\par Assuming that is easy to get $2^L-1$ ($L=\log(D)$) angle parameters $\bm{\vartheta}_{t} = (\vartheta_{t}^{0}, \cdots, \vartheta_{t}^{2^{t-1}-1} )$ ($t = 1, 2, \cdots, L$) from the updated $\mathbf{w}(n)$ after the last iteration. The angle $\vartheta_{t}^{j}$ satisfies
     \begin{equation}
     {\cos(\vartheta^{j}_{t})=\frac{h_{t}^{2j}}{h^{j}_{t-1}}}, ~~ {\sin (\vartheta^{j}_{t}) = \frac{h^{2j+1}_{t}}{h^{j}_{t-1}}},
     \label{eq:14+1}
     \end{equation}
for $t =1, \cdots, L$, where $h^{j}_{t-1} = \sqrt{( h^{2j}_{t} )^{2} + ( h^{2j+1}_{t} )^{2}}$ and $j=0, \cdots, 2^{t-1}-1$. In particular, $h^{j}_{L} = \mathbf{w}^{j}(n)$ for $j = 0, 1, \cdots, D-1$. And there are defined
    \begin{equation}
        U(\bm{\vartheta}_{t}) =
        \begin{cases}
            \sum\limits_{j = 0}^{2^{t-1}-1} {|j\rangle \langle j| \otimes R({\vartheta}^{j}_{t}) \otimes I \{ {L-t}\}},&t=2, \cdots, L \\
            R({\vartheta}^{j}_{t}) \otimes I\{ {L-t}\}, &t=1
        \end{cases},
     \label{eq:14+2}
     \end{equation}
where $I\{ {L-t}\}$ is represented as the gate $I$ applied on $(L-t)$ qubits.

\par After that, a quantum state
     \begin{equation}
     U(\bm{\vartheta}_{L}) \cdots U(\bm{\vartheta}_{2}) U(\bm{\vartheta}_{1})|0^{\otimes \log{D}}\rangle = {\sum_{j = 0}^{D-1}{c_{2}\mathbf{w}^{j}(n)\left| j \right\rangle}},
     \label{eq:14+3}
     \end{equation}
is generated in time $O(D)$ by applying the operation $U(\bm{\vartheta}_{t})$ for $t = 1, 2, \cdots, L$. Furthermore, a register in state $|1\rangle$ is appended. The overall system is in the state
     \begin{equation}
     \left| {\phi( \mathbf{w}(n))} \right\rangle = \sum_{j = 0}^{D-1}{c_{2}\mathbf{w}^{j}(n)\left| j \right\rangle}|1\rangle.
     \label{eq:14+4}
     \end{equation}
To further interpret this method, an example is given in the appendix \ref{appendix B}.

\par According to Eq. (\ref{eq:14+4}), the state in Eq. (\ref{eq:14}) can be rewritten as
     \begin{equation}
         \frac{1}{\sqrt{D}} \left[ \sum\limits_{j = 0}^{D-1} |j\rangle \sqrt{1 - ({c_{2}\mathbf{w}^{j}(n)})^{2}} |0 \rangle +  | {\phi( \mathbf{w}(n))} \rangle \right].
     \label{eq:14+5}
     \end{equation}
It means that the above two methods both allow us to extract the parameter $\mathbf{w}(n)$ information into the quantum state $\left| {\phi( \mathbf{w}(n))} \right\rangle$. On the basis of the current quantum technology, we choose the second method which is more feasible, and denote the process as $U_{\mathbf{w}}$.
\subsection{\label{sec:3.1}Local training by quantum parallel computing (QGD algorithm)}

\par  Now, we propose a QGD algorithm. It enables clients to estimate the gradient of the model in parallel based on their respective local data. According to Eq. (\ref{eq:2}), with the help of the two operations $U_{\mathbf{x}_i}$ and $U_{\mathbf{w}}$ of quantum data preparation, the process of the QGD algorithm is described as follows.

\par $(B1)$ \textit{Generate an intermediate quantum state.}
\par The task of computing the gradient involves an inner product computation, which needs $O(MD)$ in classical computers. In the era of big data, this time is costly. Here, we generate an intermediate state that contains the information of $\mathbf{x}_{i} \cdot \mathbf{w}$. This state facilitates subsequent parallel estimation.

\par $(B1.1)$ A quantum state is initialized as
     \begin{equation}
         \frac{1}{\sqrt{M}}{\sum\limits_{i = 0}^{M-1}| i \rangle_{1}}| 0^{\otimes{\log{D}}} \rangle_{2} | 0^{\otimes q}\rangle_{3} |0\rangle_{4} | 0\rangle_{5}.
     \label{eq:15}
     \end{equation}

\par $(B1.2)$ The Hadamard gate is performed on the fifth register. Then, a controlled operation $\left| i \right\rangle \left\langle i  \right| \otimes U_{\mathbf{x}_{i}} \otimes \left|  0 \right\rangle \left\langle 0 \right| + I \otimes U_{\mathbf{w}} \otimes \left| 1 \right\rangle \left\langle 1\right|$ is applied to produce a state
     \begin{equation}
         \frac{1}{\sqrt{2M}}{\sum\limits_{i = 0}^{M-1}| i \rangle_{1}}\left[{|  {\phi( \mathbf{x}_{i} )} \rangle _{24}\left|  0 \right\rangle _{5} + | {\phi( \mathbf{w}(n) )}\rangle_{24}| 1\rangle _{5}} \right].
     \label{eq:16}
     \end{equation}
\par $(B1.3)$ Subsequently, the Hadamard gate is implemented on the fifth register to get
     \begin{equation}
         \left| \psi_{1} \right\rangle = \frac{1}{\sqrt{M}}{\sum\limits_{i = 0}^{M-1}{\left| i \right\rangle_{1} \left| \Psi_{i,n} \right\rangle_{245}}},
     \label{eq:17}
     \end{equation}
where
     \begin{equation}
     \begin{split}
         &|\Psi_{i,n}\rangle =\\
          &\frac{1}{2}\left[( |  {\phi( \mathbf{x}_{i})}\rangle + | {\phi( \mathbf{w}(n))}\rangle)|0\rangle + ( |  {\phi( \mathbf{x}_{i})}\rangle -| {\phi( \mathbf{w}(n))}\rangle )|1\rangle \right].
     \end{split}
     \label{eq:17+1}
     \end{equation}
The state $|\Psi_{i,n}\rangle$ can be rewritten as
     \begin{equation}
     \begin{split}
        &| \Psi_{i,n}\rangle =\cos\theta_{i} |\psi_{i}^{0}\rangle + \sin\theta_{i} | \psi_{i}^{1}\rangle\\
        &= (\cdots)_{245}^{\bot} + \frac{1}{2\sqrt{D}}( {{\sum\limits_{j = 0}^{D-1}{c_{1}\mathbf{x}_{i}^{j}}} |j\rangle - {\sum\limits_{j^{\prime} = 0}^{D-1}{c_{2}^{\prime} \mathbf{w}^{j^{\prime}}(n)}} | j^{\prime}\rangle} )_{2} |11\rangle_{45},
     \end{split}
     \label{eq:18}
     \end{equation}
where $\theta_{i} \in \left[ 0,\frac{\pi}{2} \right]$. It is easy to verify that
     \begin{equation}
         \sin^{2}\theta_{i} = \frac{c_{1}^{2}\| \mathbf{x}_{i} \|^{2} + c_{2}^{{\prime}2}\| \mathbf{w}\|^{2} - 2c_{1}c_{2}^{\prime}(\mathbf{x}_{i} \cdot \mathbf{w})}{4D}
     \label{eq:19}
     \end{equation}
and $c_{2}^{{\prime}} = \sqrt{D}{c_{2}}$. By observing Eq. (\ref{eq:18}) and Eq. (\ref{eq:19}), it can be found that the essential information is provided by the system when its fourth and fifth registers are both in state $|1\rangle$. It means that the superposition of $|0\rangle$ also does not affect the extraction of the required information if choosing the state of Eq. (\ref{eq:14+5}). Thus, the first method (in step $(A2)$) is also suitable for our algorithm, which $c_{2}^{{\prime}} = c_{2}$.

\par \textit{$(B2)$ Calculate the $F(\mathbf{x}_{i} \cdot \mathbf{w})$ in parallel.}
\par The approximation of $F(\mathbf{x}_{i} \cdot \mathbf{w})$ should be estimated and stored in a quantum state. To achieve this goal, the $\theta_i$ is needed to estimate via quantum phase estimation which the unitary operation is defined as
     \begin{equation}
        Q_{i} = - \mathcal{A}_{i}S_{00}\mathcal{A}_{i}^{\dagger}S_{11},
     \label{eq:20}
     \end{equation}
where $\mathcal{A}_{i}\left| {0}^{\otimes{[\log{(D)}+2]}} \right\rangle = \left| \Psi_{i,n} \right\rangle$, $S_{00} = I^{\otimes {[\log{(D)}+2]}} - 2{\left|0^{\otimes{[\log{(D)}+2]}} \right\rangle\left\langle 0^{\otimes{[\log{(D)}+2]}} \right|}$ and $S_{11}=I^{\otimes{\log{(D)}}} \otimes \left( I^{\otimes{2}}- 2| 1^{\otimes{2}} \rangle \langle1^{\otimes{2}}| \right)$. Mathematically, the eigenvalues of $Q_i$ are $e^{\pm 2\mathbf{i}\theta_{i}}$ $(\mathbf{i} = \sqrt{- 1})$ and the corresponding eigenvectors are $\left| \Psi_{i,n}^{\pm} \right\rangle = \frac{1}{\sqrt{2}}\left( {\left| \psi_{i}^{0} \right\rangle \pm \mathbf{i}\left| \psi_{i}^{1} \right\rangle} \right)$), respectively. Based on the set of its eigenvectors, $\left| \Psi_{i,n} \right\rangle $ can be rewritten as $\left| \Psi_{i,n} \right\rangle = - \frac{\mathbf{i}}{\sqrt{2}}\left( {e^{\mathbf{i}\theta_{i}} \left| \Psi_{i,n}^{+} \right\rangle - e^{- \mathbf{i}\theta_{i}}\left| \Psi_{i,n}^{-} \right\rangle} \right)$. The procedure of estimating the $F(\mathbf{x}_{i} \cdot \mathbf{w})$ is displayed as follows.

\par $(B2.1)$ Performing the QPE on $Q_i$ with the state $\left| \psi_{1} \right\rangle \left| 0^{\otimes l} \right\rangle$ for some $l= O\left( \text{log}{\epsilon_{\theta}^{-1}} \right)$, an approximate state
     \begin{equation}
     \begin{split}
        &\left| \psi_{2} \right\rangle = \\
         &\frac{- \mathbf{i}}{\sqrt{2M}}{\sum\limits_{i = 0}^{M-1}\left| i \right\rangle_{1}}\left( e^{\mathbf{i}\theta_{i}}\left| \Psi_{i,n}^{+} \right\rangle \left|  {\widetilde{\theta}}_{i} \right\rangle - e^{- \mathbf{i}\theta_{i}}\left| \Psi_{i,n}^{-} \right\rangle {\left| -  {\widetilde{\theta}}_{i} \right\rangle}  \right)_{2456}
     \end{split}
     \label{eq:21}
     \end{equation}
is obtained, where ${\widetilde{\theta}}_{i} \in \mathbb{Z}_{2^{l}}$ satisfies $\left| {\theta_{i} - {\widetilde{\theta}}_{i}\pi/2^{l}} \right| \leq \epsilon_{\theta}$. Then, the quantum state
     \begin{equation}
        \left| \left. \psi_{3} \right\rangle \right. = \frac{1}{\sqrt{M}}{\sum\limits_{i = 0}^{M-1}{\left|i \right\rangle_{1}\left|\Psi_{i,n} \right\rangle_{245}| {{\sin}^{2}( {\widetilde{\theta}}_{i})} \rangle_{6}}}
     \label{eq:22}
     \end{equation}
is generated by using the sine gate. It holds for the fact that $ { {\sin}^{2} ( {\widetilde{\theta}}_{i} ) } = {{\sin}^{2}( -{\widetilde{\theta}}_{i})}$.

\par $(B2.2)$ According to Eq. (\ref{eq:19}), it is needed to access $\left\| \mathbf{x}_{i} \right\|$ to compute $\mathbf{x}_{i} \cdot \mathbf{w}$. Combining with the operation $U_{nf}$ and the quantum arithmetic operations \cite{zhou2017}, we can get
     \begin{equation}
        \left|\psi_{4} \right\rangle = \frac{1}{\sqrt{M}}{\sum\limits_{i = 0}^{M-1}\left| i \right\rangle_{1}}\left| \Psi_{i,n} \right\rangle_{245}\left|{\mathbf{x}_{i} \cdot \mathbf{w}} \right\rangle_{6}\left| \left\| \mathbf{x}_{i} \right\| \right\rangle_{7}\left|\left\| \mathbf{w} \right\| \right\rangle_{8}.
     \label{eq:23}
     \end{equation}

\par $(B2.3)$ An oracle $O_F$ is supposed to achieve any function which has a convergent Taylor series \cite{cong2016}. Combining with $O_y$, the function $F(*)$ could be implemented (a simple example is described in Sec. \ref{sec:5} ). The state becomes
     \begin{equation}
        \left|\psi_{5} \right\rangle = \frac{1}{\sqrt{M}}{\sum\limits_{i = 1}^{M}\left| i \right\rangle_{1}}\left| \Psi_{i,n} \right\rangle_{245}\left|F({\mathbf{x}_{i} \cdot \mathbf{w}}) \right\rangle_{6}\left| \left\| \mathbf{x}_{i} \right\| \right\rangle_{7}\left|\left\| \mathbf{w} \right\| \right\rangle_{8}.
     \label{eq:24}
     \end{equation}
\par $(B2.4)$ Next, a register in the state $|0\rangle$ is appended as the last register and rotated it to $\left| \phi_{i} \right\rangle  = c_{3}F\left( {\mathbf{x}_{i} \cdot \mathbf{w}} \right)\left|0 \right\rangle + \sqrt{1 - \left( c_{3}F\left( {\mathbf{x}_{i} \cdot \mathbf{w}} \right) \right)^{2}}\left| 1 \right\rangle$ in a controlled manner, where $c_{3} = {1/{\max\limits_{i}\left| {F\left( {\mathbf{x}_{i} \cdot \mathbf{w}} \right)} \right|}}$. This results in the overall state
     \begin{equation}
     \begin{split}
        &\left|\psi_{6} \right\rangle = \\
        &\frac{1}{\sqrt{M}}{\sum\limits_{i = 0}^{M-1}\left| i \right\rangle_{1}}\left| \Psi_{i,n} \right\rangle_{245}\left|F({\mathbf{x}_{i} \cdot \mathbf{w}}) \right\rangle_{6}\left| \left\| \mathbf{x}_{i} \right\| \right\rangle_{7}\left|\left\| \mathbf{w} \right\| \right\rangle_{8} {\left| \phi_{i} \right\rangle_{9}}.
     \end{split}
     \label{eq:25}
     \end{equation}
\par
\begin{figure*}[!t]
        \centering
        \includegraphics[height=0.26\textwidth,width=0.92\textwidth]{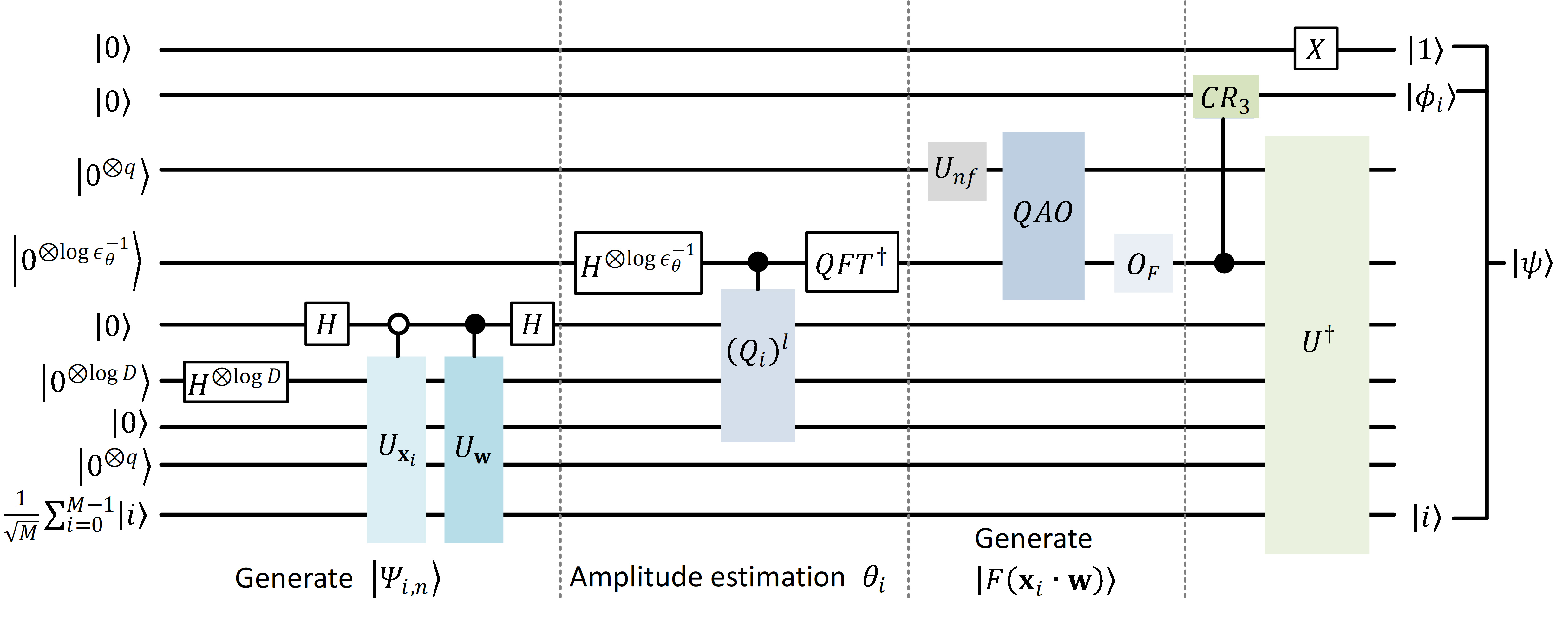}
        \caption{Quantum circuit diagram of $\mathcal{A}_{\psi}$. Here, $q$ is the number of qubits required to adequately store the data information and $\epsilon_\theta$ is the tolerance error for estimating $\theta_{i}$. Furthermore, $QAO$ denotes the quantum arithmetic operations, $CR_3$ represents the controlled rotation operation about $|\phi_{i}\rangle$ and $U^{\dagger}$ labels the inverse operations in the step $(B2.5)$.}
    \label{fig:3}
    \end{figure*}

 $(B2.5)$ The inverse operations of steps $(B1.2)-(B2.3)$ are performed on $\left|\psi_{6} \right\rangle$. Afterwards, a register in the state $|1\rangle$ is added to obtain
     \begin{equation}
        \left|\psi \right\rangle = \frac{1}{\sqrt{M}}{\sum\limits_{i = 0}^{M-1}{\left| i \right\rangle _{1}\left| \phi_{i} \right\rangle_{9}}}\left|1 \right\rangle_{10}.
     \label{eq:26}
     \end{equation}
For convenience, the $\mathcal{A}_{\psi}$ is marked as the operations which achieve $\mathcal{A}_{\psi}|00\cdots 0\rangle = |\psi\rangle$. Its schematic quantum circuit is given in Fig. \ref{fig:3}.

\par \textit{$(B3)$ Estimate the gradient $\bm{g}^{j}\left( {\mathbf{w}(n)} \right)$ with swap test.}
\par $(B3.1)$ Three registers in state $ \frac{1}{\sqrt{M}}{\sum_{i = 0}^{M-1}{\left|i \right\rangle_{1} \left|j \right\rangle_{2} }}\left| 0 \right\rangle_{3} $ are prepared. Performing $O_{X}$ on it to generate the state
     \begin{equation}
        \frac{1}{\sqrt{M}}{\sum\limits_{i = 0}^{M-1}{ | i \rangle_{1} |j \rangle_{2}}} | \mathbf{x}_{i}^{j} \rangle_{3}.
     \label{eq:27}
     \end{equation}
\par $(B3.2)$ The controlled rotation operation ($|0\rangle \rightarrow \sqrt{1 - ({c_{1}\mathbf{x}_{i}^{j}})^{2}} \left|0 \right\rangle + c_{1}\mathbf{x}_{i}^{j}\left|1 \right\rangle$) is implemented to get
     \begin{equation}
        \frac{1}{\sqrt{M}}{\sum\limits_{i = 0}^{M-1}{ |i \rangle_{1} \left|j \right\rangle_{2} }} | \mathbf{x}_{i}^{j} \rangle_{3} \left[ {\sqrt{1 - ( {c_{1} {\mathbf{x}}_{i}^{j}} )^{2}}\left|0 \right\rangle + c_{1} \mathbf{x}_{i}^{j} |1 \rangle} \right]_4.
     \label{eq:28}
     \end{equation}

\par $(B3.3)$ The inverse operation of $O_X$ is performed. After that, we can obtain the state
     \begin{equation}
        | \chi^{j} \rangle = \frac{1}{\sqrt{M}}{\sum\limits_{i = 0}^{M-1}{| i \rangle_{1} | 0 \rangle_{3} }}\left[ {\sqrt{1 - ( {c_{1} \mathbf{x}_{i}^{j}} )^{2}} |0 \rangle + c_{1}\mathbf{x}_{i}^{j} | 1 \rangle} \right]_{4},
     \label{eq:29}
     \end{equation}
via undoing the register $|j\rangle$.
\par $(B3.4)$ In order to obtain the gradient, the technology of swap test \cite{buhrman2001} is utilized. Combining the processes of generating the states $\left| \psi \right\rangle$ and $\left| \chi^{j} \right\rangle$, a quantum state $\frac{1}{\sqrt{2}}\left( \left|0 \right\rangle \left|\psi \right\rangle + \left|1 \right\rangle \left| \chi^{j} \right\rangle \right)$ can be constructed. Then, measuring the first register to see whether it is in the state $\left| + \right\rangle = \frac{1}{\sqrt{2}}\left( \left| 0 \right\rangle + \left|1 \right\rangle\right)$. The measurement has the success probability
     \begin{equation}
        P = \frac{1}{2} + \frac{1}{2}\left\langle \psi \middle| \chi^{j} \right\rangle.
     \label{eq:30}
     \end{equation}
According to Eq. (\ref{eq:2}), the $\bm{g}^{j}\left( {\mathbf{w}(n)} \right) = {(2P - 1)}/{({c_{1}c_{3}})}$ can be calculated. Hence, it is possible for ${\rm Bob}_k$ to obtain the local gradient $\bm{g}_{k}\left( \mathbf{w} \right) = \left( {\bm{g}_{k}^{0}\left( {\mathbf{w}(n)} \right),\bm{g}_{k}^{1}\left( {\mathbf{w}(n)} \right), \cdots, \bm{g}_{k}^{D-1} \left( {\mathbf{w}(n)} \right)} \right)^{T}$ by repeating the steps of the above algorithm with his data.
\subsection{\label{sec:3.2} Model aggregation with QSMC protocol and update}
\par We will design a protocol to safely compute the federated gradients $\bm{G}\left( \mathbf{w} \right) = {\sum_{k = 1}^{K}{\beta_{k}}{\bm{g}_{k}\left( \mathbf{w} \right)}}$ in this section. That is, calculating $\bm{G}\left( \mathbf{w} \right)$ without revealing the local gradient ${\bm{g}_{k}\left( \mathbf{w} \right)}$. To do it, the server Alice is assumed to be semi-honest who may misbehave on her own but cannot conspire with others. Moreover, the federated gradients are needed to be accurate to ${\gamma}^{-1}$. This means that $\gamma {\beta_{k}} {\bm{g}_{k}^{j} ( \mathbf{w} )} \geq 0$. Simply, the $\gamma {\beta_{k}} {\bm{g}_{k}^{j} ( \mathbf{w} )}$ is marked as $\mu_{k}^{j}$. And supposing that ${\sum_{k = 1}^{K} \mu_{k}^{j}} < S$. The further details are described as follows.

\par $(C1)$ \textit{Preparation for multi-party quantum communication.}

\par Alice announces ${\gamma}$ and the global dataset scale $(\sum_{k = 1}^{K}{M_{k}})$. At same time, the participants (server and clients) choose $m$ numbers $d_i$ ($i=1,2,\cdots,m$) which are mutually prime and satisfy $d_{1} \times d_{2} \times \cdots \times d_{m} = S$. Subsequently, ${\rm Bob}_{k}$ $(k=1,2,\cdots,K)$ calculates his secret
     \begin{equation}
        s_{k,i}^{j} =  \mu_{k}^{j} ~\text{mod}~d_{i}.
     \label{eq:31}
     \end{equation}
Alice produces a $d_{i}$-level $(K+1)$-particle GHZ state
     \begin{equation}
        \left| \Psi \right\rangle = \frac{1}{\sqrt{d_{i}}}{\sum\limits_{q = 0}^{d_{i} - 1} |  q \rangle ^{\bigotimes(K + 1)}},
     \label{eq:32}
     \end{equation}
and marks the $(K+1)$ particles by $Q_{0},Q_{1},\cdots,Q_{K}$.

\par $(C2)$ \textit{Distribution of quantum pairs.}
\par For the sake of checking the presence of eavesdroppers, Alice prepares $K$ sets of $\delta$ decoy states, where each decoy photon randomly is in one of the states from the set $V_{1} = \left\{ \left| p \right\rangle \right\}_{p = 0}^{d_{i} - 1}$ and $V_{2} = \left\{ QFT\left|p \right\rangle\right\}_{p = 0}^{d_{i} - 1}$, where $QFT$ is represented as the quantum Fourier transform \cite{nielsen2010quantum}. These sets are denoted as $E_{1},E_{2},\cdots,E_{K}$, respectively. Then Alice inserts $Q_k$ into $E_k$ at a random position, and sends them to ${\rm Bob}_k$ for $k=1,2,\cdots,K$.

\par $(C3)$ \textit{Security checking of quantum channel.}
\par After receiving $\delta+1$ particles, ${\rm Bob}_k$ sends acknowledgements to Alice. Subsequently, the positions and the bases of the decoy photons are announced to ${\rm Bob}_k$ by Alice. ${\rm Bob}_k$ measures the decoy photons and returns the measurement results to Alice who then calculates the error rate by comparing the measurement results with initial states. If the error rate is higher than the threshold determined by the channel noise, Alice cancels this protocol and restarts it. Otherwise, the protocol is continued.

\par $(C4)$ \textit{Measurement of particles and encoding of transmission information.}
\par ${\rm Bob}_k$ extracts all the decoy photons and discards them. Then, server and clients perform a measurement $\left\{ QFT\left|  p \right\rangle \right\}_{p = 0}^{d_{i} - 1}$ on the remaining particles, respectively. The measurement results record as $o_{s,i},o_{1,i},\cdots,o_{K,i}$ and these satisfy $o_{s,i}+o_{1,i}+\cdots+o_{K,i}=0 ~\text{mod}~ d_{i}$. Subsequently, ${\rm Bob}_k$ encodes his data $s_{k,i}^{\prime j} = s_{k,i}^{j} + o_{k,i}$ and sends it to Alice.

\par $(C5)$ \textit{Computation of federated gradient by server.}
\par At this stage, Alice accumulates all the results $s_{k,i}^{\prime j}$ to compute
     \begin{equation}
     \begin{split}
        &\left( {o_{s} + o_{1,i} + s_{1,i}^{j} + o_{2,i} + s_{2,i}^{j}\cdots + o_{K,i} + s_{K,i}^{j}} \right)~\text{mod}~d_{i}\\
        =&\left( {\mu_{1}^{j}} + {\mu_{2}^{j}} + \cdots + {\mu_{K}^{j}} \right)~\text{mod}~d_{i}.
     \end{split}
     \label{eq:33}
     \end{equation}
For  $i=1,2,\cdots,m$, Alice can obtain $m$ equations such as Eq. (\ref{eq:33}). According to the Chinese remainder theorem, Alice compute the summation
     \begin{equation}
        \left( {\sum\limits_{k = 1}^{K} {\mu_{k}^{j}}} \right)~\text{mod}~S = ~{\sum\limits_{k = 1}^{K} {\mu_{k}^{j}}}.
     \label{eq:34}
     \end{equation}
And it is easy to get the federated gradient
     \begin{equation}
        \bm{G}^{j}\left( \mathbf{w} \right) = \frac{1}{\gamma}\sum\limits_{k = 1}^{K} {\mu_{k}^{j}} = { \sum\limits_{k = 1}^{K} {\beta_{k}} {\bm{g}_{k}^{j} ( \mathbf{w} )} }.
     \label{eq:35}
     \end{equation}

\par After the similar processes, the federated gradient $( \bm{G}^{0}( \mathbf{w} ), \bm{G}^{1}( \mathbf{w} ), \cdots, \bm{G}^{D-1}( \mathbf{w} ) )$ could be obtained by Alice. And she updates the global model parameters $\mathbf{w}( {n + 1}) = \mathbf{w}(n) - \alpha \times \bm{G}\left( {\mathbf{w}(n)} \right)$. In order to exhibit the process of model aggregation more clearly, a concrete example is presented in the appendix \ref{appendix C}.
\subsection{\label{sec:3.3} Model evaluation and distribution via classical communication networks}
\par The server (Alice) needs to evaluate whether the model should be further optimized after one round of training in QFLGD. Similar to classical FL, Alice utilizes the smoothness of the gradient to evaluate the model performance. Specifically, the server sends a termination training signal and announces the global parameters when ${\sum\limits_{j = 0}^{D-1}\left[ {\bm{G}^{j}\left( \mathbf{w}(n) \right)} \right]^{2}} \leq \varepsilon$. Otherwise, she distributes the updated model parameters to clients for new training.
\section{\label{sec:4}Analysis}
\par In this section, we provide a brief analysis of the proposed framework. As discussed previously, the QGD algorithm (shown in Sec. \ref{sec:3.1}) enables clients to accelerate the training gradients on a local quantum computer. The QSMC protocol (shown in Sec. \ref{sec:3.2}) gives a method to securely update the federated parameters to protect the privacy of clients' data. Therefore, two main aspects are considered in the analysis. One is the time complexity of local training (the QGD algorithm). The other is the security of model aggregation (the QSMC protocol).
\subsection{\label{sec:4.1}Time complexity of local training (the QGD algorithm)}
\par  In the QFLGD framework, assuming that $M_{1} \leq M_{2} \leq \cdots \leq M_{K} \leq M$. Namely, the dataset scale is at most $M$. And all clients need to accomplish the gradient training before calculating the federated gradient. Thus the waiting time for the distributed training gradient is the time consumed to train the dataset which scale is $M$. In the following, the time complexity of the QGD algorithm is analyzed with the $M$-scale dataset.
\par In the data preparation period (the Sec. \ref{sec:3.0}), the time consumption is caused by the processes of $U_{\mathbf{x}_{i}}$ and $U_{\mathbf{w}}$, which generate the states $\left| {\phi( \mathbf{x}_{i})} \right\rangle $ and $\left| {\phi( \mathbf{w}(n))} \right\rangle $ about data information. It could be implemented in time $O(\text{polylog}(MD)+D+q)$ with the help of the $O_{X}$, $U(\bm{\vartheta}_{t})$ and the controlled rotation operation \cite{HHL,wossnig2018}. The $q$ is represented as the number of qubits which store the data information. Afterwards, $U_{\mathbf{x}_{i}}$ and $U_\mathbf{w}$ are applied to produce the state $\left| \psi_{1} \right\rangle$ in step $(B1)$ of local training (the Sec. \ref{sec:3.1}). Hence, step $(B1)$ can be implemented in time $O(\text{polylog}(MD)+D+q)$.
\par In step $(B2)$, we first consider the complexity of the unitary operation $\mathcal{A}_{i}$. It contains $H$, $U_{\mathbf{x}_{i}}$, and $U_\mathbf{w}$ which take time $O(\text{polylog}(MD)+D+q)$. Then, the QPE block needs $O\left( {1/\epsilon_{\theta}} \right)$ applications of $Q_{i}$ to estimate the $\theta_{i}$ within error $\epsilon_{\theta}$ \cite{brassard2002}. Therefore, the time complexity of step $(B2.1)$ is $O[(\text{polylog}(MD)+D+q)/{\epsilon_{\theta}}]$. The runtime $O( \text{log}( {1/\epsilon_{\theta}}))$ \cite{zhou2017} of implementing the sine gate can be ignored, which is much smaller than the QPE.
\par Next, the time complexity of step $(B2.2)$ and step $(B2.3)$ are discussed. The main operation of the two steps includes $U_{nf}$, $O_{y}$, and the quantum arithmetic operation, which are performed to calculate $| F ( {\mathbf{x}} \cdot {\mathbf{w}} ) \rangle $ in time $O[{(\text{polylog}(D))/{\epsilon_{m}}} +\text{polylog}(M)+ q]$. In step $(B2.4)$, the time complexity of the controlled rotation is $O(q)$. Step $(B2.5)$ takes time $O[ {(\text{polylog}(D))/{\epsilon_{m}}} + { ( \text{polylog}(MD) + D + q ) / { \epsilon_{\theta} } } ]$ to implement the inverse operations of steps $(B1.2)$-$(B2.3)$. Putting all the steps together to get the time complexity of step $(B2)$ as $O[ {(\text{polylog}(D))/{\epsilon_{m}}} + { ( \text{polylog}(MD) + D + q ) / { \epsilon_{\theta} } } ]$.
\par In step $(B3)$, the processes of generating the $| \chi^{j} \rangle$ (described in steps $(B3.1)$-$(B3.3)$) are accomplished in time $O(\text{polylog}(MD)+q)$. According to step $(B2)$, a copy of the quantum state $| \psi \rangle$ is produced in time $O[{(\text{polylog}(D))/{\epsilon_{m}}} + {(\text{polylog}(MD)+D+q)/{\epsilon_{\theta}}}]$. The swap test is applied $O\left( {{P\left( {1 - P} \right)}/\epsilon_{P}^{2}} \right) = O\left( {1/\epsilon_{P}^{2}} \right)$ times to get the result $P$ within error $\epsilon_{P}$ in step $(B3.4)$ \cite{rebentrost2014}. And each swap test should prepare a copy of $| \chi^{j} \rangle$ and $| \psi \rangle$. Therefore, the runtime is $\left\{ [{(\text{polylog}(D))/{\epsilon_{m}}} + {(\text{polylog}(MD)+D+q)/{\epsilon_{\theta}}}] \epsilon_{P}^{-2} \right\}$ in step $(B3)$, that is the complexity of obtaining the desired result.
\par For convenience, we assume that $\mathbf{w}^{j}$, $\mathbf{x}_{i}^{j}=O(1)$, then $\|  \mathbf{w} \|$, ${\max\limits_{i}\left\| \mathbf{x}_{i} \right\|} = O( \sqrt{D})$. Therefore, $q={\text{polylog}(D)}$ could fulfill the number of qubit required to store data information. In addition, taking $\epsilon_{m}$, $\epsilon_{\theta}$, and $\epsilon_{P}$ equaling to $\epsilon$. After that, the complexity of the entire quantum algorithm to get $\bm{g}^{j}\left( \mathbf{w} \right)$ $(j=0,1,\cdots,D-1)$ in each iteration can further simplify into
     \begin{equation}
        O\left\{ {D\left[ {\left( \text{polylog}\left( {MD} \right) + D \right)/\epsilon^{3}} \right]} \right\}.
     \label{eq:36}
     \end{equation}
This means that the time complexity of training gradient is $O(D^{2} \text{polylog}(MD))$ when $\epsilon^{-1} = \text{log}(MD)$, achieving exponential acceleration on the number of data samples. Furthermore, the elements of $\mathbf{w}$ can also be accessed in time $O( \text{polylog}(D) )$ if they are timely writing in QRAM. In this case, the proposed algorithm has exponential acceleration on the number $M$ and the quadratic speedup in the dimensionality $D$, compared with the classical algorithm whose runtime is $O(M D^{2})$.

\subsection{\label{sec:4.2}Security analysis of model aggregation (the QSMC protocol)}
\par In this section, the security of model aggregation (the QSMC protocol) will be analyzed. For the secure multi-party computing, the attacks from outside and all participants are the challenges, which have to deal with. In the following, we will show these attacks are invalid to our protocol.

\par Firstly, the outside attacks are discussed. In this protocol, the decoy photons is used to prevent the eavesdropping. This idea is derived from the BB84 protocol \cite{bennett1984}, which has been proved unconditionally safe. Here, we take the intercept-resend attack as an example to demonstrate. If an outside eavesdropper Eve attempts to intercept the particles sent from Alice and replaces them with his own fake particles, he will introduce extra error rate $1 - \left( \frac{d_{i}+1}{2d_{i}} \right)^{\delta}$. Therefore, Eve will be detected in step $(C3)$ through security checking analysis.

\par Secondly, the participant attacks are analyzed. In the proposed protocol, the participants include the server (Alice) and clients (${\rm Bob}_{k}$, $k=1,2,\cdots, K$) who can access more information. Therefore, the participant attacks from dishonest clients or server should be considered.

\par For the participant attack from dishonest clients, only the extreme case of $K-1$ clients ${\rm Bob}_{1}, \cdots, {\rm Bob}_{k-1}, {\rm Bob}_{k+1}, \cdots, {\rm Bob}_{K}$ colluding to steal the secret from ${\rm Bob}_{k}$ is considered here, because $K-1$ clients have the most powerful strength. In this case, even if the dishonest clients share their information, they cannot deduce $o_{k}$ without the help of Alice. That means they cannot obtain the secret $s_{k,i}^{j}$ by $s_{k,i}^{\prime j} = s_{k,i}^{j} + o_{k}$. Thus, our algorithm can resist the collusion attack of dishonest ${\rm Bob}_{k}$.

\par For the attack from Alice, the semi-honest Alice may steal the private information of ${\rm Bob}_{k}$ without conspiring with any one. In step (C4), Alice collects $s_{k,i}^{\prime j}$ for $k=1,2,\cdots,K$. However, she still cannot learn $s_{k,i}^{j}$ due to the lack of knowledge about $o_{k}$ which from clients.

\section{\label{sec:5}Application: Training the Federated Linear Regression Model}

\subsection{\label{sec:5.1}Quantum federated linear regression algorithm}
\par Linear regression (LR) is an important supervised learning algorithm, which establishes a model of the relationship between the variable $\mathbf{x}_{i}$ and the observation $y_i$. It has wide application in the scientific fields of biology, finance, and so on \cite{geron2022}. LR models are also usually fitted by minimizing the function in Eq. (\ref{eq:1}) and choosing ${f\left( {\mathbf{x}_{i} \cdot \mathbf{w}} \right)} = {\mathbf{x}_{i} \cdot \mathbf{w}} + {b}$ ($b$ is a migration parameter).

\par In this section, we apply the QFLGD framework to train the LR model. In the training process, we need to implement the function
     \begin{equation}
        F\left( {\mathbf{x}_{i} \cdot \mathbf{w}} \right) = \mathbf{x}_{i} \cdot \mathbf{w} + b - y_{i}.
     \label{eq:38}
     \end{equation}
The state $| {\mathbf{x}_{i} \cdot \mathbf{w}} \rangle$ about ${\mathbf{x}_{i} \cdot \mathbf{w}}$ can be generated according to the QGD. Then, the state $| \mathbf{x}_{i} \cdot \mathbf{w} + b - y_{i} \rangle$ is produced in the following steps.

\par (S1) The oracle $O_y$ is applied on the state ${| {\mathbf{x}_{i} \cdot \mathbf{w}} \rangle}_{A} |0\rangle_{B}^{\otimes{q}}$ to get
     \begin{equation}
        {| {\mathbf{x}_{i} \cdot \mathbf{w}} \rangle}_{A} |y_{i}\rangle_{B}^{\otimes{q}},
     \label{eq:39}
     \end{equation}
in time $O(\text{log}(M))$.

\par (S2) After obtaining ${| {\mathbf{x}_{i} \cdot \mathbf{w}} \rangle} = {| e_{q},e_{q-1},\cdots,e_{1} \rangle}$ and ${|y_{i}\rangle} = {|t_{q},t_{q-1},\cdots,t_{1} \rangle}$, we implement the $QFT$ on $| {\mathbf{x}_{i} \cdot \mathbf{w}} \rangle_{A}$ to result in
     \begin{equation}
        \left[ {\left|{\phi_{1}(e)} \right\rangle\otimes \left|{\phi_{2}(e)} \right\rangle\otimes \cdots \otimes \left| {\phi_{q}(e)} \right\rangle} \right]_{A}\left|y_{i} \right\rangle_{B},
     \label{eq:40}
     \end{equation}
where $\left|{\phi_{j}(e)} \right\rangle = \frac{1}{\sqrt{2}}\left( \left| 0 \right\rangle + e^{2\pi \mathbf{i}0.e_{j}e_{j - 1}\cdots e_{1}}\left|1 \right\rangle\right)$ for $j=1,2,\cdots,q$.

\par (S3) Subsequently, the controlled rotation operation ${I \otimes |0\rangle \langle 0 |} + { R_{j^{\prime}} \otimes |1\rangle \langle 1 |}$ ($j^{\prime}=1,\cdots,{j}$) are performed on $\left|{\phi_{j}(e)} \right\rangle$ and $|t_{q-j+1}\rangle$ ($j=1,2,\cdots,q$), we can get
     \begin{equation}
        \left[ {\left|{\phi_{1}(e-t)} \right\rangle\otimes \left|{\phi_{2}(e-t)} \right\rangle\otimes \cdots \otimes \left| {\phi_{q}(e-t)} \right\rangle} \right]_{A}\left|y_{i} \right\rangle_{B},
     \label{eq:41}
     \end{equation}
where the $R_{j^{\prime}}$ is defined as $|0\rangle \langle 0 | + e^{{- 2\pi \mathbf{i}}/2^{j^{\prime}}}|1\rangle \langle 1 |$ and $\left|{\phi_{j}(e-j)} \right\rangle = \frac{1}{\sqrt{2}}\left( | 0 \rangle + e^{2\pi \mathbf{i}(0.e_{j}\cdots e_{1}-0.t_{j}\cdots t_{1})} |1 \rangle\right)$.

\par
\begin{figure*}[!t]
        \centering
        \includegraphics[height=0.3\textwidth,width=0.92\textwidth]{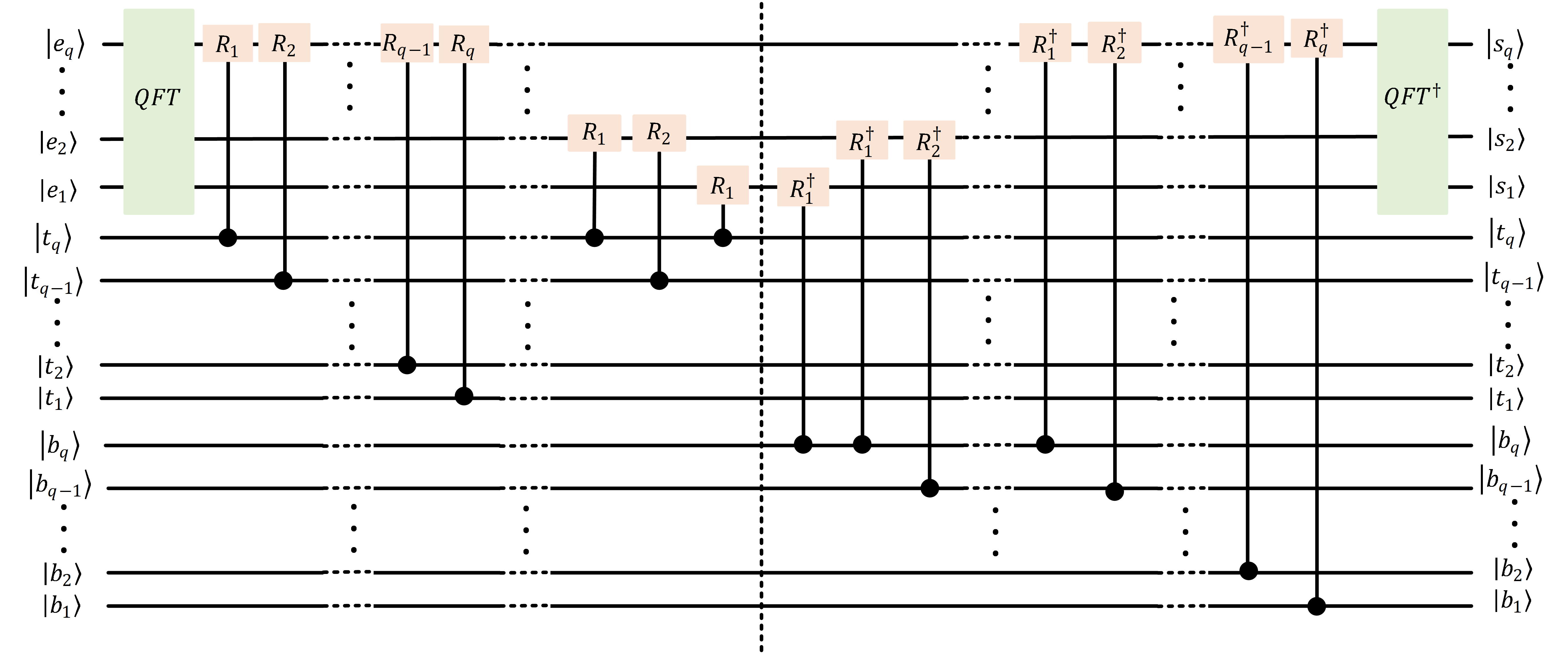}
        \caption{Quantum circuit of the quantum computing $| \mathbf{x}_{i} \cdot \mathbf{w}  - y_{i} + b\rangle$. The left side of the dotted line is the circuit of subtraction, and the right side is the circuit of addition. It is worth noting that the $QFT^{\dagger}QFT = I$, so we omitted them left and right of the dotted line.}
    \label{fig:4}
    \end{figure*}

 (S4) Inverse $QFT$ is applied on the register $A$, the state becomes
     \begin{equation}
        \left[ {\left|{e_{q} - t_{q}} \right\rangle\otimes \left|{e_{q-1} - t_{q-1}} \right\rangle\otimes \cdots \otimes \left| {e_{1} - t_{1}} \right\rangle} \right]_{A}\left|y_{i} \right\rangle_{B}.
     \label{eq:42}
     \end{equation}
Thus, the state $| \mathbf{x}_{i} \cdot \mathbf{w} - y_{i} \rangle$ can be obtained from the register A. Similarly, we can implement addition by changing the operation of step (S3) to ${I \otimes |0\rangle \langle 0 |} + { R_{j^{\prime}}^{\dagger} \otimes |1\rangle \langle 1 |}$. Thus, the state $| \mathbf{x}_{i} \cdot \mathbf{w}  - y_{i} + b\rangle$ could be obtained and its quantum circuit is presented in Fig. \ref{fig:4}. The operations of these processes are labeled as $U_{s}$, which are implemented in time $O( \text{log}(M) + q^{2})$. Combining with the QFLGD framework, the quantum federated linear regression (QFLR) model could be constructed by algorithm \ref{algorithm2}.

\begin{algorithm}[]
    \renewcommand{\thealgocf}{1}
    \caption{Quantum federated LR algorithm}\label{algorithm2}
	\KwIn{The variate $\mathbf{X} \in \mathbb{R}^{M \times D}$, the observed vector $\mathbf{y} \in \mathbb{R}^{M}$, the initial parameter  $\mathbf{w}(0) \in \mathbb{R}^{D}$, the migration parameter $b$, the learning rate $\alpha$ and the preset values $c_{i}$ $(i=1,2,3)$;}
	\KwOut{The parameter $\mathbf{w}$ and the model $y = \mathbf{w}^{T}\mathbf{x}_{i} + b$;}
	\For {${\sum\limits_{j = 0}^{D-1}\left( {\bm{G}^{j}\left( {\mathbf{w}(n - 1)} \right)} \right)^{2}}$ $>$ $\varepsilon$}
    {
    \hangafter 1
    \hangindent 2.5em
        \qquad Step 1: $K$ clients prepare quantum states with their sensitive data according to the methods in Sec. \ref{sec:3.0};\\
    \hangafter 1
    \hangindent 2.5em
        \qquad Step 2: The clients apply the QGD Algorithm (in Sec. \ref{sec:3.1}) and $U_s$ to local training $\bm{g}_{k}\left( {\mathbf{w}(n)} \right)$ ($k=1,2,\cdots,K$) respectively; \\
    \hangafter 1
    \hangindent 2.5em
        \qquad Step 3: The clients and the server use the QSMC protocol (in Sec. \ref{sec:3.2}) to secure calculate the federated gradient $\bm{G}\left( \mathbf{w}(n) \right) = \left( \bm{G}^{0}\left( {\mathbf{w}(n)} \right), \bm{G}^{1}\left( {\mathbf{w}(n)} \right),\cdots,\bm{G}^{D-1}\left( {\mathbf{w}(n)} \right) \right)$, and the server updates global model parameter $\mathbf{w}( {n + 1}) = \mathbf{w}(n) - \alpha \times \bm{G}\left( {\mathbf{w}(n)} \right)$; \\
    \hangafter 1
    \hangindent 2.5em
        \qquad Step 4: The server evaluates model performance and sends the updated model parameter to $K$ clients;\\
	}
\end{algorithm}

\subsection{\label{sec:5.2}Numerical Simulation}

\par In this section, the numerical simulation of the QFLR algorithm will be presented. In our simulation, two clients (${\rm Bob}_{1}$, ${\rm Bob}_{2}$) trained the QFLR model with a sever (Alice). The experiment is implemented on the IBM Qiskit simulator. The initial weight $\mathbf{w}(0) = (0.866,0.5)$ and the migration parameter $b=0$ are selected by Alice. ${\rm Bob}_{1}$ chooses an input vector $\mathbf{x}=(2,3.464)$ which corresponds the observation $y=2.464$. Another client ${\rm Bob}_{2}$ selects an input vector $\mathbf{x}^{\prime}=(2.5,4.33)$ and the corresponding observation $y^{\prime}=2.33$.

\par In the process of training the federated linear regression model, the main is to achieve the perfect $F\left( {\mathbf{x}_{i} \cdot \mathbf{w}} \right)$ calculation. That is, quantum computing $F\left( {\mathbf{x}_{i} \cdot \mathbf{w}} \right)$ values are required to be able to be stored in quantum registers with small error. An experiment of this step is presented with the data of ${\rm Bob}_{1}$. For convenience, setting $c_1=1/4$, $c_2=1$, and the error $\epsilon_{\theta} = 0.0001$ of quantum phase estimation. By substituting these into Eq. (\ref{eq:38}), the result $F\left( {\mathbf{x} \cdot \mathbf{w}} \right) = 1$ could be obtained. It can also be computed by
     \begin{equation}
     \begin{split}
        F\left( {\mathbf{x} \cdot \mathbf{w}} \right) = 4 - 16{\sin}^{2}\left( {\widetilde{\theta}\pi/2^{4}} \right) + 0 - 2.464,
     \end{split}
     \label{eq:43}
     \end{equation}
according to Eq. (\ref{eq:19}).

\par
\begin{figure*}[htb]
        \centering
        \subfloat[]{
            \includegraphics[height=0.16\textwidth,width=0.75\textwidth]{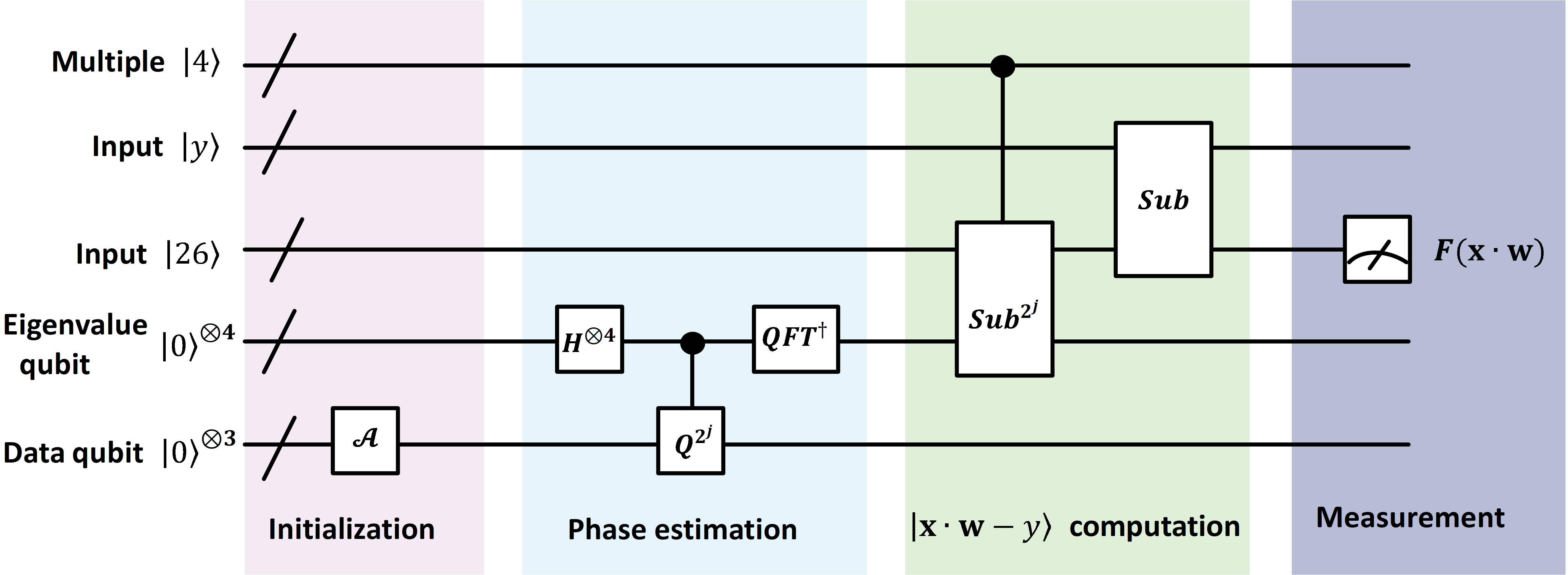}
        \label{fig:5a}
        }
        \newline
        \subfloat[]{
            \includegraphics[height=0.10\textwidth,width=0.30\textwidth]{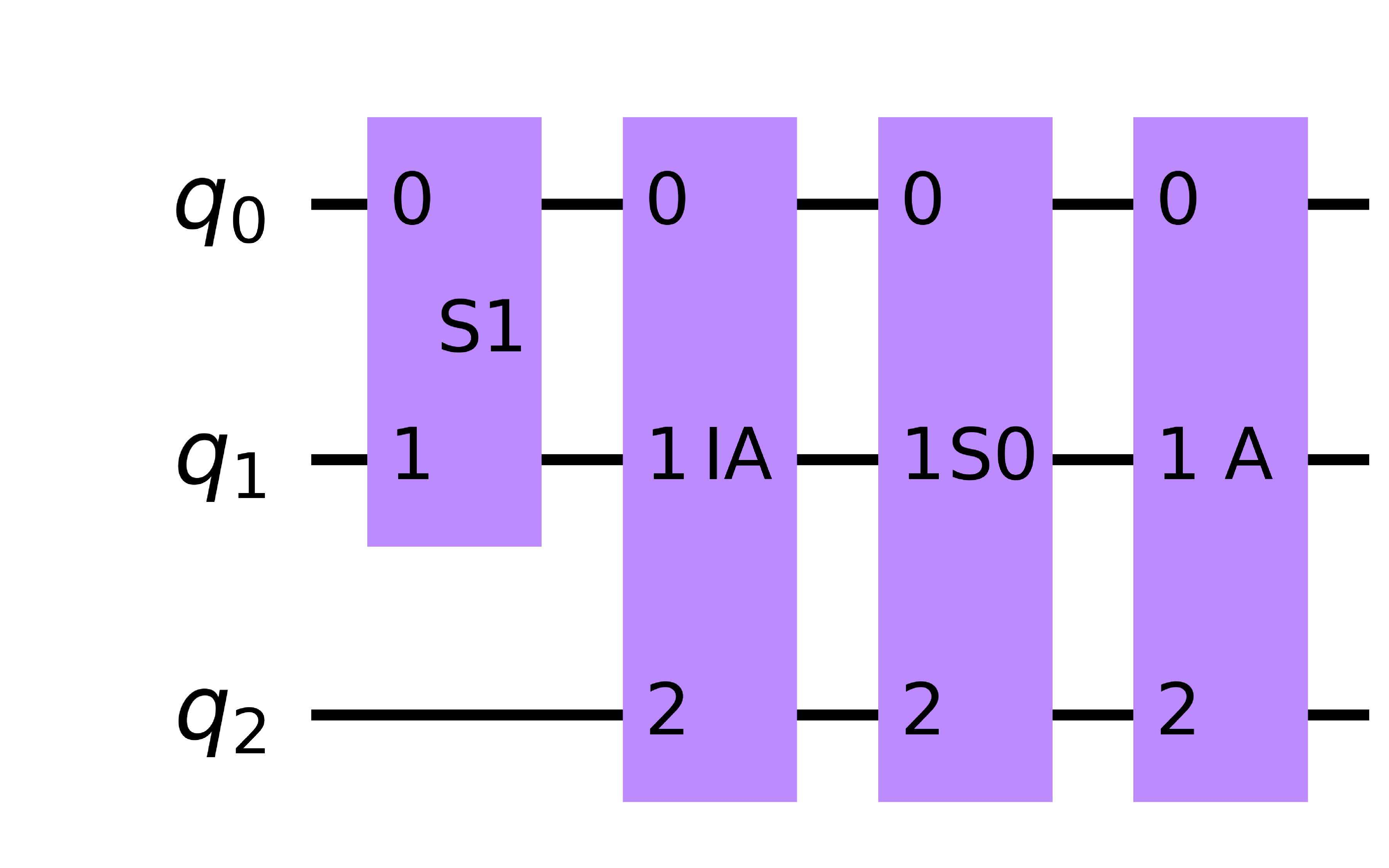}
        \label{fig:b0}
        }
        \hfil
        \subfloat[]{
            \includegraphics[height=0.10\textwidth,width=0.30\textwidth]{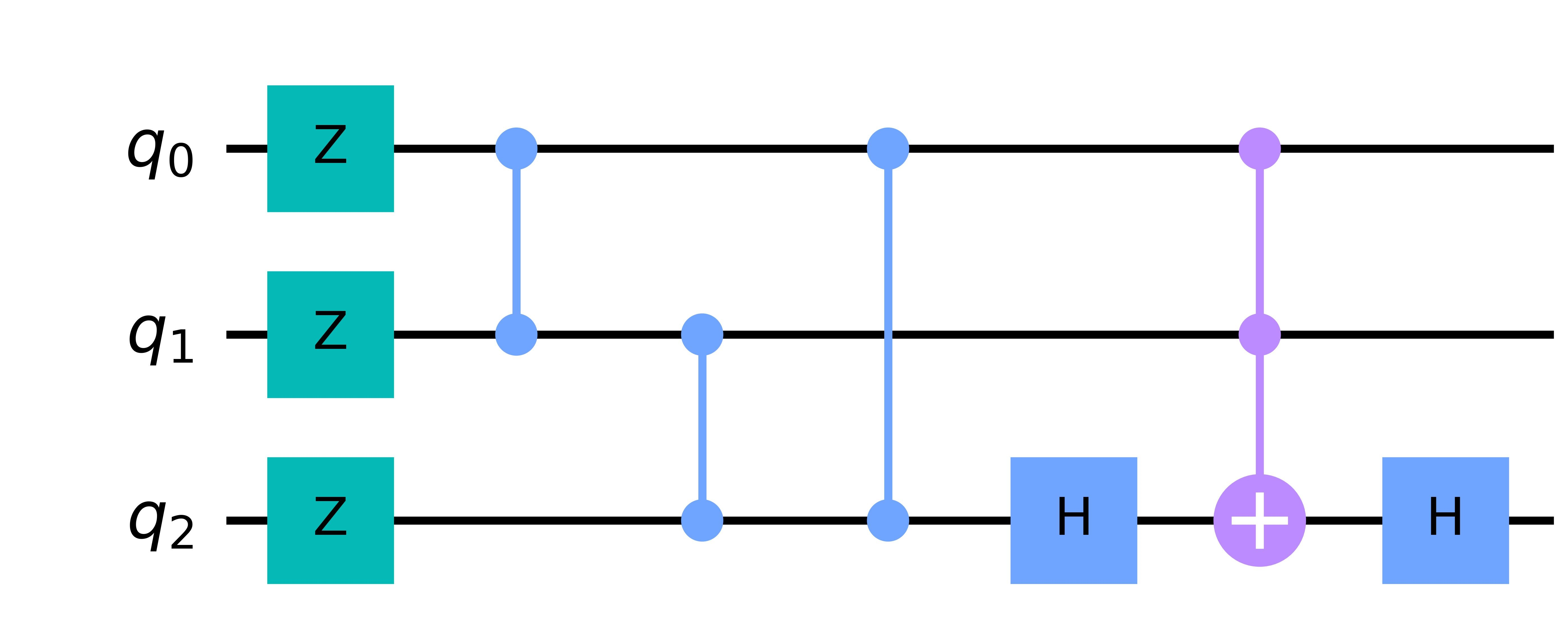}
        \label{fig:b1}
        }
        \hfil
        \subfloat[]{
            \includegraphics[height=0.08\textwidth,width=0.20\textwidth]{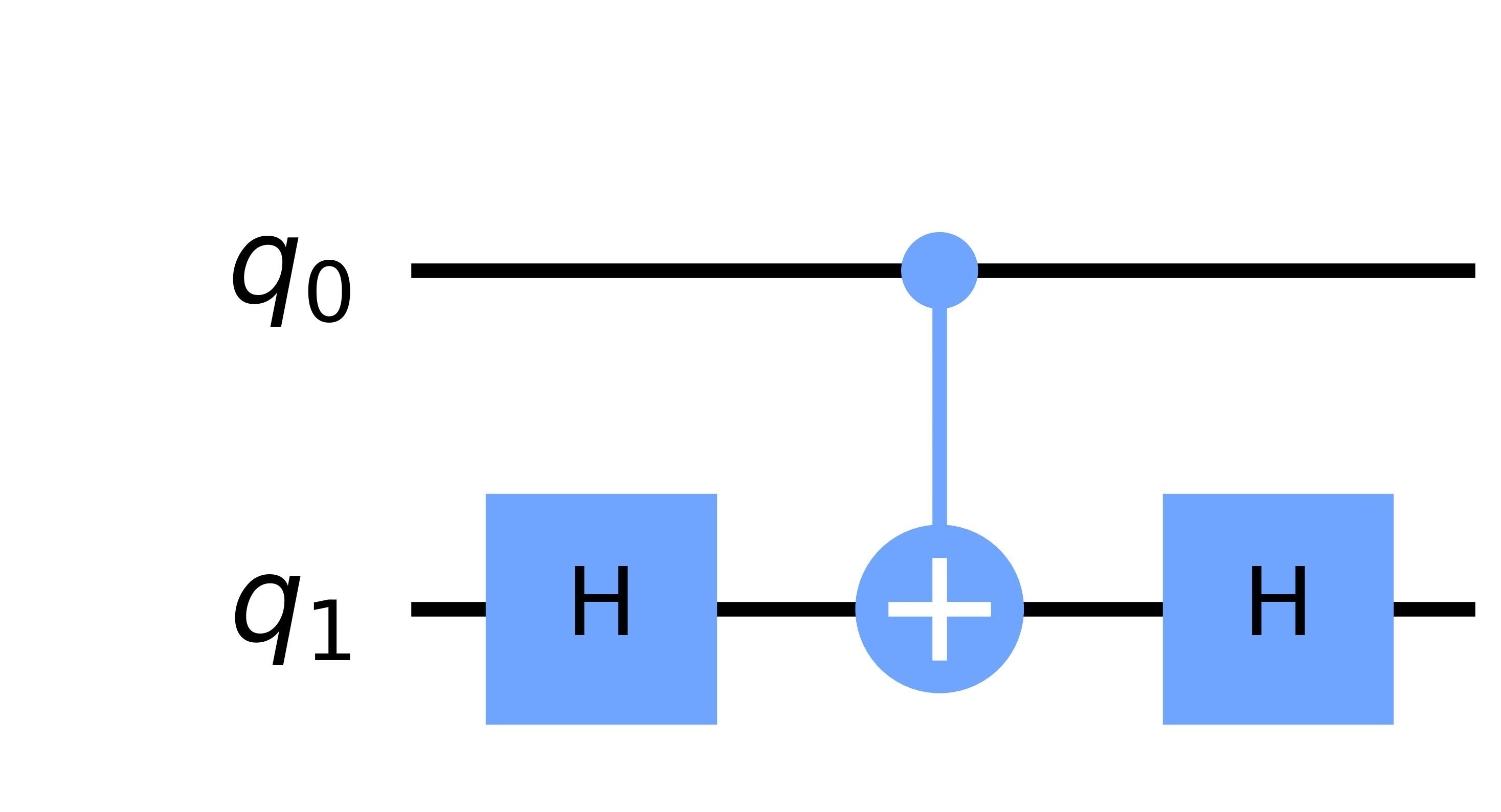}
        \label{fig:b2}
        }
        \newline
        \subfloat[]{
            \includegraphics[height=0.25\textwidth,width=0.88\textwidth]{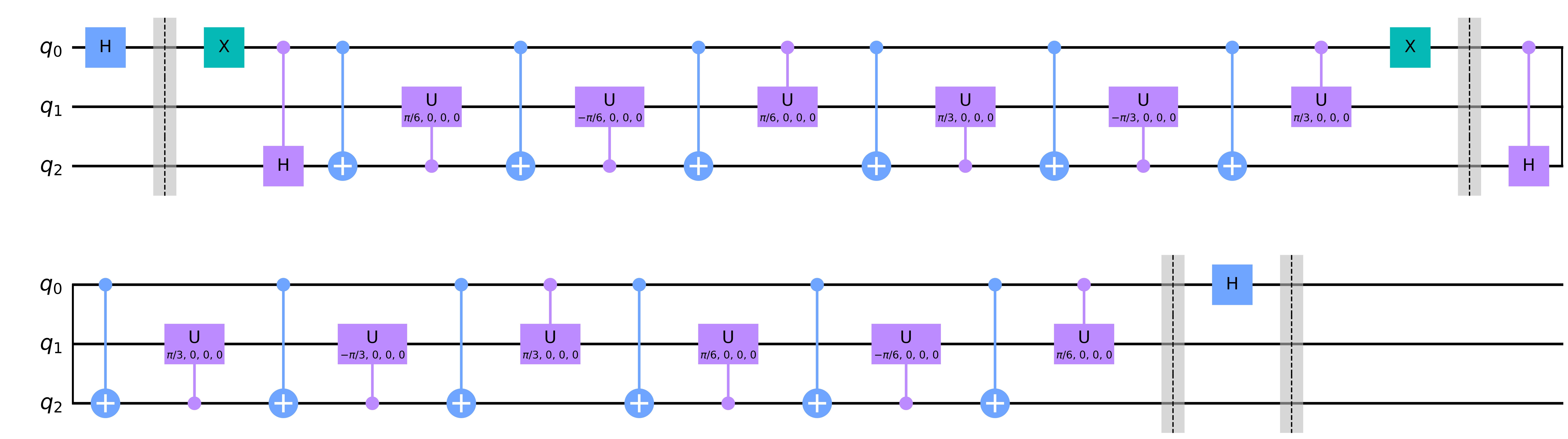}
        \label{fig:5c}
        }
        \caption{The construction of the gates for solving $ {\mathbf{x}_{i} \cdot \mathbf{w}}$. (a) The optimized circuit for solving Eq. (\ref{eq:43}). (b) The circuit of the operation $Q$. (c) The construction of $2|000\rangle\langle000|-I$. (d) The circuit of the operation $I-2|11\rangle \langle11|$.  (e) The construction of $\mathcal{A}$.}
        \label{fig:5}
    \end{figure*}

\begin{figure*}[htb]
        \centering
        \subfloat[]{
            \includegraphics[height=0.22\textwidth,width=0.32\textwidth]{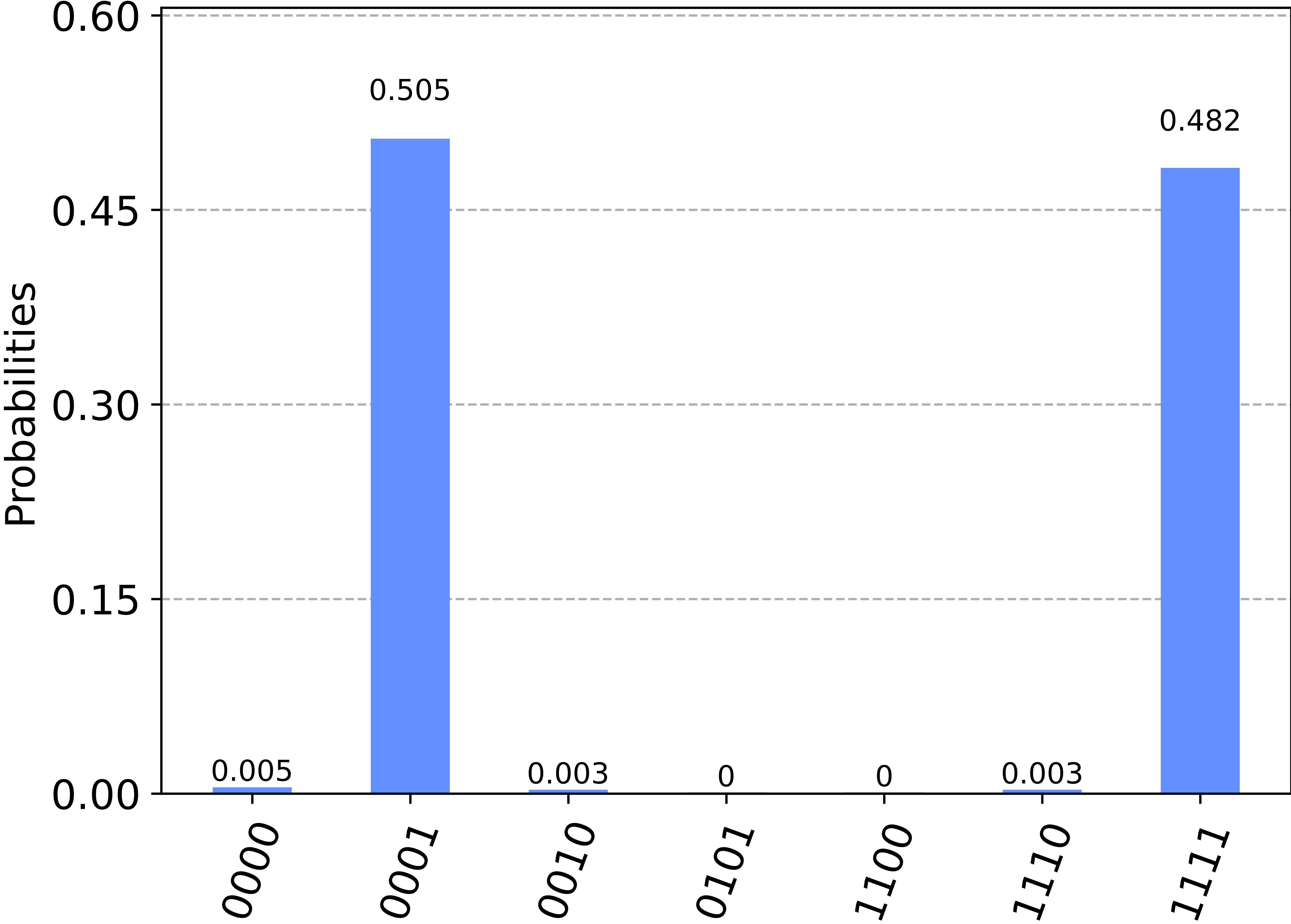}
        \label{fig:6a}
        }
        \hfil
        \subfloat[]{
            \includegraphics[height=0.22\textwidth,width=0.32\textwidth]{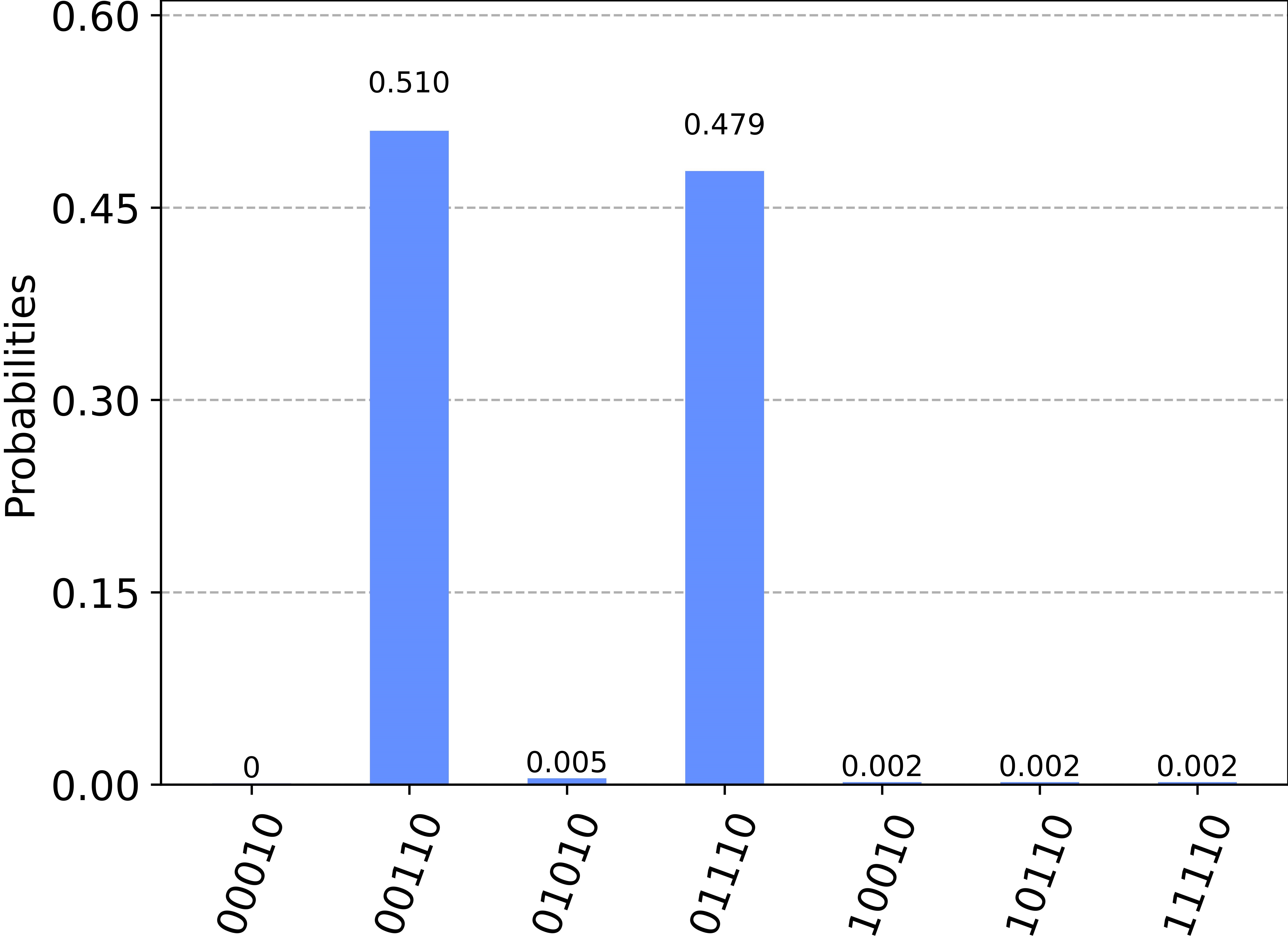}
        \label{fig:6b}
        }
        \caption{Experimental results. (a) The resulting diagram of the eigenvalue registers measurement for QPE. These two significant values correspond to $e^{\mathbf{i}{\theta}}$ and $e^{-\mathbf{i}{\theta}}$. (b) The proportional results of the estimation of $ {\mathbf{x} \cdot \mathbf{w}}$.}
        \label{fig:6}
    \end{figure*}

 With the fact of ${\sin}^{2}( {\widetilde{\theta}\pi/2^{4}} ) \approx \widetilde{\theta}^{2}/26$ and the most probable result $| 0001 \rangle$ (see Fig. \ref{fig:6}(a)) from the QPE, Eq. (\ref{eq:43}) can be rewritten as
     \begin{equation}
        6.5 \times F\left( {\mathbf{x} \cdot \mathbf{w}} \right) \approx 26 - 4\widetilde{\theta} - 16.
     \label{eq:44}
     \end{equation}
Its circuit is designed as exhibited and encoded via Qiskit (see Fig. \ref{fig:5}). In Fig. \ref{fig:5}(e), the matrix form of $U(\gamma,\phi,\lambda)$ is
     \begin{equation}
        U\left( {\gamma,\phi,\lambda} \right) =
        \begin{bmatrix}
        {\cos \left( {\gamma/2} \right)} & {- e^{\mathbf{i}\lambda}\sin \left( {\gamma/2} \right)} \\
        {e^{\mathbf{i}\phi}\sin \left( {\gamma/2} \right)} & {e^{\mathbf{i}(\phi + \lambda)}\cos \left( {\gamma/2} \right)} \\
        \end{bmatrix}.
     \label{eq:45}
     \end{equation}
With the help of the IBM's simulator (aer$\underline{~}$simulator), the measurement results can be obtained which are shown in Fig. \ref{fig:6}(b). From Fig. \ref{fig:6}(b), two values ($| 00110 \rangle$ and $| 01110 \rangle$) stand out, which have a much higher probability of measurement than the rest. Based on the analysis of the phase estimation results, selecting result $| 00110 \rangle$ with a high probability of $0.510$. It means $F\left( {\mathbf{x} \cdot \mathbf{w}} \right)\approx 0.923$. Compared with the theoretical result (shown in Eq. (\ref{eq:43})), the experimental result has an error of $0.077$ which is tolerable. Subsequently, ${\rm Bob}_{1}$ can estimate $\bm{g}_{1}^{1} \approx 1.846 $ and $\bm{g}_{1}^{2} \approx 3.197$ by performing swap test. At same time, ${\rm Bob}_{2}$ estimates $F\left( {\mathbf{x}^{\prime} \cdot \mathbf{w}} \right)\approx 2.115$, $\bm{g}_{2}^{1} \approx 3.197$, and $\bm{g}_{2}^{2} \approx 9.157$ of his data via similar experiment.

\par As the analogous process of the example shown in the appendix \ref{appendix C}, Alice calculates the federated gradient $G=(3.57,6.18)$ via the QSMC protocol. Theoretical analysis shows that the error is within $2\%$ of the actual solution $(3.5,6.06)$ which is obtained in the example. Thus, the training algorithm is found to be successful.

\section{\label{sec:6}Conclusions}
\par This work focuses on the design of the QFLGD for distributed quantum networks that can securely implement FL over an exponentially large data set. We first gave two methods of quantum data preparation, which can extract static data information and dynamic parameter information into logarithmic qubits. Then, we put forth the QGD algorithm to allow the time-consuming gradient calculation to be done on a quantum computer. In this way, the clients can estimate some urgently needed results of gradient training in parallel based on quantum superposition. The time complexity analysis is shown that our algorithm is exponentially faster than its classical counterpart on the number of data samples when the error $1/\epsilon=\text{log}(MD)$. Furthermore, the QGD algorithm could also achieve quadratic speedup on the dimensionality of the data sample if the parameters $\mathbf{w}$ are stored in QRAM timely. And, we proposed a QSMC protocol to calculate the federated gradient securely. The evidence is demonstrated that the proposed protocol could resist some common outside and participant attacks, such as the intercept-resend attack. Finally, we indicated how to apply it to train a federated linear regression model and simulated some steps with the help of the IBM Qiskit simulator. The results also showed the effectiveness of QFLGD. In summary, the presented framework demonstrates the intriguing potential of achieving large-scale private distributed learning with quantum technologies and provides a valuable guide for exploring quantum advantages in real-life machine learning applications from the security perspective.

\par We hope the proposed framework can further be realized on a quantum platform with the gradual maturity of quantum technology. For example, how to implement the whole QFLGD process on the noisy intermediate-scale quantum (NISQ) devices is worth further exploration, and we will make more efforts.

\section*{Acknowledgments}
This work was supported by National Natural Science Foundation of China (Grants No. 62171131, 61976053, and 61772134), Fujian Province Natural Science Foundation (Grant No. 2022J01186 and 2023J01533), and Innovation Program for Quantum Science and Technology (Grant No. 2021ZD0302901).


{\appendices
\section{\label{appendix A}Implement the Unitary Operation $U_{nf}$}
\par In this appendix, we describe the implementation of a unitary operation $U_{nf}$, which could generate a state about the $2$-norm of $\mathbf{x}_{i}$. Its steps as shown in the following.
\par (1) A quantum state is initialized as
     \begin{equation}
        \left|\varphi_{1} \right\rangle = \frac{1}{\sqrt{D}}{\sum\limits_{j = 0}^{D-1}\left| i \right\rangle_{1}}\left| j \right\rangle _{2}\left|0 \right\rangle_{3}.
     \label{eq:46}
     \end{equation}

\par (2) The oracle $O_X$ is performed to obtain
     \begin{equation}
        \left|\varphi_{2} \right\rangle = \frac{1}{\sqrt{D}}{\sum\limits_{j = 0}^{D-1}\left| i \right\rangle_{1}}\left|j \right\rangle _{2}\left|\mathbf{x}_{i}^{j} \right\rangle_{3}.
     \label{eq:47}
     \end{equation}

\par (3) A register in the state $|0\rangle$ is appended as the last register and rotated to $\sqrt{1 - (c_{1}\mathbf{x}_{i}^{j})^{2}}\left| 0 \right\rangle + c_{1}\mathbf{x}_{i}^{j}\left| 1 \right\rangle$. After that, the system becomes
     \begin{equation}
        |\varphi_{3}\rangle = \frac{1}{\sqrt{D}}{\sum\limits_{j = 0}^{D-1} |i\rangle _{1}} |j\rangle_{2} |\mathbf{x}_{i}^{j}\rangle_{3}\left[ {\sqrt{1 - ( c_{1}\mathbf{x}_{i}^{j})^{2}} |0\rangle + c_{1}\mathbf{x}_{i}^{j} | 1\rangle} \right]_{4},
     \label{eq:48}
     \end{equation}
where $c_{1} = {1/{\max_{i,j}| \mathbf{x}_{i}^{j}|}}$. We can observe the ancilla register in the state $|1\rangle$ with probability $P_{1} = {{c_{1}^{2}\left\| \mathbf{x}_{i} \right\|^{2}}/D}$. The state $\left|\varphi_{3} \right\rangle$ can be rewritten as
     \begin{equation}
        \left| i \right\rangle_{1}\otimes\left( {\sqrt{1 - P_{1}}\left| g \right\rangle \left| 0 \right\rangle + \sqrt{P_{1}}\left|  a \right\rangle \left|  1 \right\rangle } \right)_{234},
     \label{eq:49}
     \end{equation}
where
     \begin{equation}
     \left| g \right\rangle = {\sum\limits_{j = 0}^{D-1}{\sqrt{\frac{1 - \left( c_{1}\mathbf{x}_{i}^{j} \right)^{2}}{D - c_{1}^{2}\left\| \mathbf{x}_{i} \right\|^{2}}}\left| j \right\rangle}}\left| \mathbf{x}_{i}^{j} \right\rangle
     \label{eq:50}
     \end{equation}
and
     \begin{equation}
     \left| a \right\rangle = \frac{1}{\left\| \mathbf{x}_{i} \right\|}{\sum\limits_{j = 0}^{D-1}{\mathbf{x}_{i}^{j}\left| j \right\rangle}}\left| \mathbf{x}_{i}^{j} \right\rangle.
     \label{eq:51}
     \end{equation}

\par (4) Appending a register in state $\left| 0 \right\rangle^{\otimes \log(\epsilon_{m}^{- 1})}$. Then, the quantum phase estimation of $- U\left( \varphi_{3} \right)S_{0}U^{\dagger}\left( \varphi_{3} \right)S_{1}$ is performed to obtain
     \begin{equation}
     \left| \varphi_{4} \right\rangle = \left|  i \right\rangle_{1}\otimes\left( {\sqrt{1 - P_{1}}\left|g \right\rangle \left|  0 \right\rangle + \sqrt{P_{1}}\left|  a \right\rangle \left| 1 \right\rangle } \right)_{234} \otimes \left| \left\| \mathbf{x}_{i} \right\| \right\rangle_{5},
     \label{eq:52}
     \end{equation}
with the help of the square root circuit \cite{bhaskar2015}. We denote the $\epsilon_{m}$ is a tolerance error of QPE, $U\left( \varphi_{3} \right)\left|  0 \right\rangle_{1234} = \left| \varphi_{3} \right\rangle$, $S_{0} = I_{1234} - 2\left| 0 \right\rangle _{1234}\left\langle 0 \right|_{1234}$ and $S_{1} = I_{123}\otimes\left( I - 2\left| 0 \right\rangle\left\langle 0 \right| \right)_{4}$.

\par (5) The inverse operations of steps (2)-(3) are applied to generated the state
     \begin{equation}
     \left| \varphi_{5} \right\rangle = \left|i \right\rangle \left| \left\| \mathbf{x}_{i} \right\| \right\rangle.
     \label{eq:53}
     \end{equation}
Therefore, the $U_{nf}:\left| i \right\rangle \left| 0 \right\rangle\rightarrow\left|  i \right\rangle \left| \left\| \mathbf{x}_{i} \right\| \right\rangle$ could be implemented in the above steps. And its running time is mainly caused by the quantum phase estimation in step (4), which takes time $O\left( {{\text{polylog}(D)}/\epsilon_{m}} \right)$. Moreover, $\left\| \mathbf{w} \right\|$ could be estimated similarly.

\section{\label{appendix B}An example of extracting \\the parameter $\mathbf{w}(n)$ information}

\par In Sec. \ref{sec:3.0}, a way to prepare a quantum state of $\mathbf{w}$ without the help of QRAM is shown in step $(A2)$. To further demonstrate it, an example is given in this appendix.

\par For convenience, supposing that $\mathbf{w}(n) = \mathbf{w} \in \mathbb{R}^{4}$. Then,we can get angle parameters $\vartheta^{0}_{1}$, $\vartheta^{0}_{2}$, and $\vartheta^{1}_{2}$ which are satisfied
     \begin{equation}
     \begin{split}
        &{\cos(\vartheta^{0}_{1})=\frac{h^{0}_{1}}{h^{0}_{0}}}, ~~~~ {\sin (\vartheta^{0}_{1}) = \frac{h^{1}_{1}}{h^{0}_{0}}},\\
        &{\cos(\vartheta^{0}_{2})=\frac{h^{0}_{2}}{h^{0}_{1}}}, ~~~~ {\sin (\vartheta^{0}_{2}) = \frac{h^{1}_{2}}{h^{0}_{1}}},\\
        &{\cos(\vartheta^{1}_{2})=\frac{h^{2}_{2}}{h^{1}_{1}}}, ~~~~ {\sin (\vartheta^{1}_{2}) = \frac{h^{3}_{2}}{h^{1}_{1}}},
     \end{split}
     \label{eq:APB1}
     \end{equation}
according to Eq. (\ref{eq:14+1}). The values of $h^{j}_{t}$ ($t=0,1,2$) are shown in Fig. \ref{fig:example}, such as $h^{j}_{2} = \mathbf{w}^{j}$ for $j=0,1,2,3$.
    \begin{figure}
        \centering
        \includegraphics[width=0.4\textwidth]{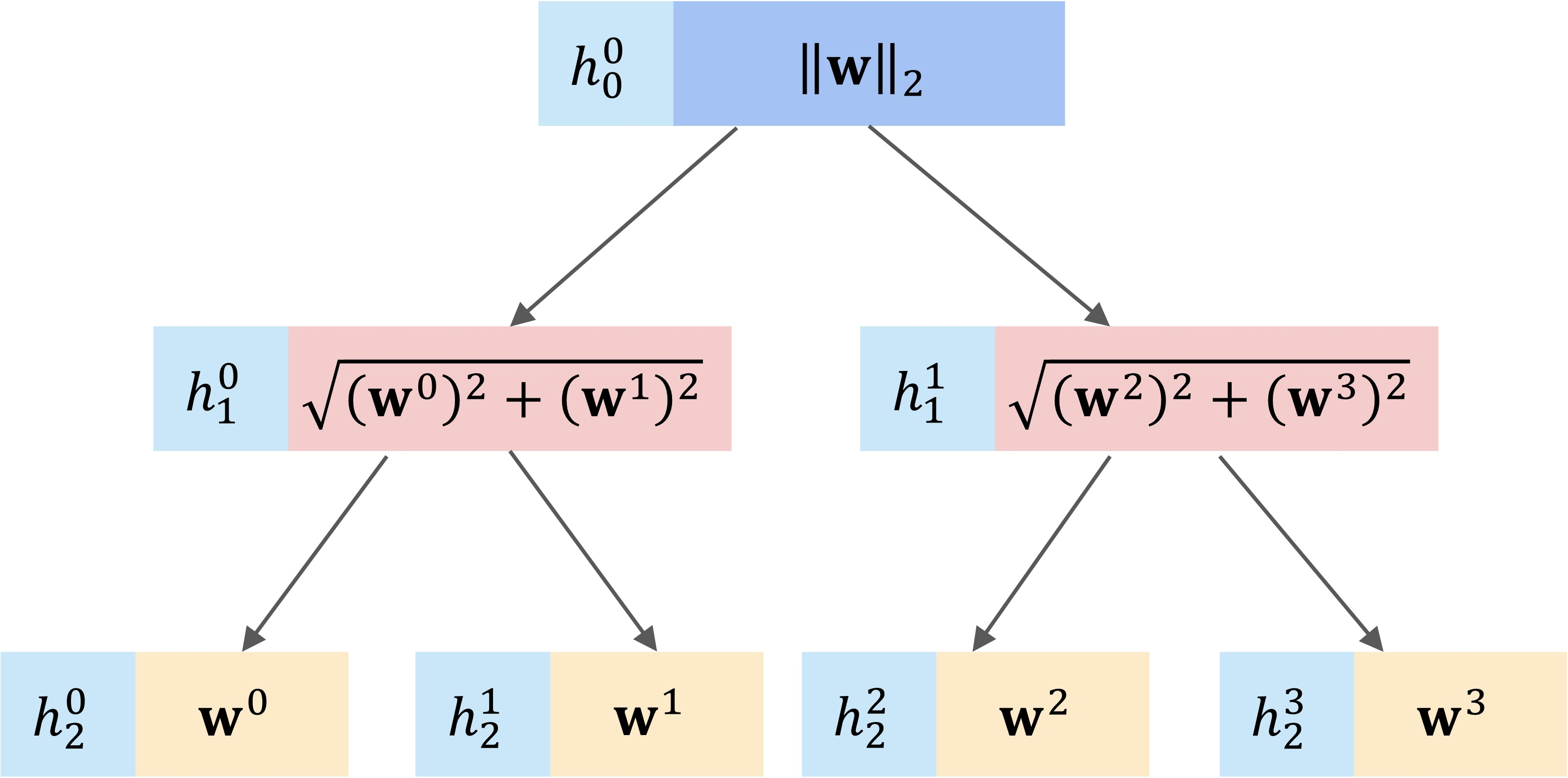}
        \caption{The data structure. $h^{j}_{t-1} = \sqrt{( h^{2j}_{t} )^{2} + ( h^{2j+1}_{t} )^{2}}$ for $t = 1,2$ and $j=0, \cdots, 2^{t-1}-1$. Moreover, $h^{j}_{2} = \mathbf{w}^{j}$ for $j = 0, 1, 2, 3$.}
    \label{fig:example}
    \end{figure}
Then, the operations are defined as
    \begin{equation}
    \begin{split}
        & U(\bm{\vartheta}_{1}) = R({\vartheta}^{0}_{1}) \otimes I, \\
        & U(\bm{\vartheta}_{2}) = |0\rangle \langle 0| \otimes R({\vartheta}^{0}_{2}) + |1\rangle \langle 1| \otimes R({\vartheta}^{1}_{2}),
    \end{split}
     \label{eq:APB2}
     \end{equation}
based on $R(\vartheta) = \begin{bmatrix} \cos{(\vartheta)} &- \sin{(\vartheta)} \\ \sin{(\vartheta)} &\cos{(\vartheta)} \\ \end{bmatrix}$. It is easy to verify that
     \begin{equation}
     U(\bm{\vartheta}_{2})U(\bm{\vartheta}_{1})|00\rangle = \frac{1}{\left\| \mathbf{w} \right\|} (\mathbf{w}^{0}|00\rangle + \mathbf{w}^{1}|01\rangle + \mathbf{w}^{2}|10\rangle + \mathbf{w}^{3}|11\rangle).
     \label{eq:APB3}
     \end{equation}
Thus, the quantum state of $\mathbf{w}$ can be obtained.

\section{\label{appendix C}An example of the model aggregation}
\par In this appendix, an example is presented to exhibit the model aggregation. Considering the model is trained by two clients (${\rm Bob}_{1}$, ${\rm Bob_{2}}$) who respectively have a $1$-scale dataset, with the help of a server (Alice). The gradients $g_{1}^{1} = 2$, $g_{1}^{2} = 3.46$, $g_{2}^{1} = 5$, and $g_{2}^{2} = 8.66$ are assumed to be gained in the QGD algorithm. Simply, the eavesdropping check phase is ignored.

\par Firstly, Alice announces the accuracy of parameters is $\gamma^{-1} = 1/100$ and the global dataset scale is $M_{1} + M_{2} = 2$. She chooses $d_{1} = 23$ and $d_{2} = 29$ with clients. After that, ${\rm Bob}_{1}$ calculates his secret
     \begin{equation}
     \begin{split}
        s_{1,1}^{1} &= ~ \mu_{1}^{1} \mod ~ 23\\
        &= 8,
     \end{split}
     \label{eq:35+1}
     \end{equation}
$s_{1,2}^{1}=13$, $s_{1,1}^{2}=12$, and $s_{1,2}^{2}=28$. At same time, ${\rm Bob}_{2}$ can get $s_{2,1}^{1} = 20$, $s_{2,2}^{1} = 18$, $s_{2,2}^{1} = 19$, and $ s_{2,2}^{2} = 27$.
\par Secondly, Alice prepares a $23$-level-$3$ particle GHZ state $\left| \Psi \right\rangle = \frac{1}{\sqrt{23}}{\sum\limits_{q = 0}^{22} |q\rangle  |q\rangle |q\rangle}$ for $d_{1} = 23$ and gives a particle to each client respectively. Then these participants perform the measurement to get $o_{s,1} = 7$, $o_{1,1} = 6$, and $o_{2,1} = 10$. ${\rm Bob}_{1}$ (${\rm Bob}_{2}$) encodes his secret by using  $o_{1,1}$ ($o_{2,1}$) and sets to Alice. The result
     \begin{equation}
     \begin{split}
        &(7+8+6+20+10)\mod~23 \\
        =&(\mu_{1}^{1}+\mu_{2}^{1})\mod~23=5,
     \end{split}
     \label{eq:35+2}
     \end{equation}
could be computed by Alice.
\par Finally, the equations
     \begin{equation}
     \begin{split}
        &(\mu_{1}^{1}+\mu_{2}^{1})\mod~23=5,\\
        &(\mu_{1}^{1}+\mu_{2}^{1})\mod~29=2,
     \end{split}
     \label{eq:35+3}
     \end{equation}
and
     \begin{equation}
     \begin{split}
        &(\mu_{1}^{2}+\mu_{2}^{2})~\mod~23=8,\\
        &(\mu_{1}^{2}+\mu_{2}^{2})~\mod~29=26,
     \end{split}
     \label{eq:35+4}
     \end{equation}
could be obtained through a similar procedure. According to the Chinese remainder theorem, the federated gradient (3.5,6.06) is easy to get.
}

\bibliography{QFLGD}
\bibliographystyle{IEEEtran}








\end{document}